\documentclass[aps,prd,showpacs,nofootinbib,superscriptaddress,preprint,tightenlines,eqsecnum,longbibliography]{revtex4}
\usepackage{amssymb,amsmath}
\usepackage{graphicx,enumerate}

\usepackage{subfigure}

\usepackage{leftidx}

\usepackage{xcolor}

\newcommand{\vd}[2]{\frac{\delta #1}{\delta #2}}   
\newcommand{\Par}[2]{\frac{\partial #1}{\partial #2}}   

\newcommand{\md}{{\mathrm d}}
\newcommand{\ed}{\md \!\!\!\! \md\, }

\newcommand{\ct}{c_{(\ell)}}

\newcommand{\pbi}[1]{{\underset{^\leftarrow}{{#1}}}} 
\newcommand{\WIHeq}{\overset{_\Delta}{=}}           

\newcommand{\A}{\mathcal{A}}
\newcommand{\sdA}{\leftidx{^+}{\mathcal{A}}}          
\newcommand{\sdF}{\leftidx{^+}{\mathcal{F}}}       

\newcommand{\W}{\mathbb{W}}
\newcommand{\V}{\mathbb{V}}
\newcommand{\U}{\mathbb{U}}
\newcommand{\Y}{\mathbb{Y}}

\newcommand{\om}{{\rm o}}   

\newcommand{\dwedge}{\wedge \!\!\!\!\!\:\! \wedge\,}

\newcommand{\be}{\begin{equation}}
\newcommand{\ee}{\end{equation}}

\begin{document}

\title{Weakly Isolated Horizons: 
	$3+1$ decomposition and canonical formulations  in self-dual variables}
\author{Alejandro Corichi}
\email{corichi@matmor.unam.mx}
\affiliation{Centro de Ciencias Matem\'aticas, Universidad Nacional Aut\'onoma de
M\'exico, UNAM-Campus Morelia, A. Postal 61-3, Morelia, Michoac\'an 58090,
Mexico}
\author{Juan D. Reyes}
\email{jdreyes@uach.mx}
\affiliation{Facultad de Ingenier\'\i a,
Universidad Aut\'onoma de Chihuahua, 
Nuevo Campus Universitario, Chihuahua 31125, Mexico}

\author{Tatjana Vuka\v{s}inac}
\email{tatjana@umich.mx}
\affiliation{Facultad de Ingenier\'ia Civil, Universidad Michoacana de San Nicol\'as de Hidalgo, 
Morelia, Michoac\'an 58000, Mexico}

\begin{abstract}
The notion of Isolated Horizons has played an important role in gravitational physics, being useful from the characterization of the endpoint of black hole mergers to (quantum) black hole entropy. 
	In particular, the definition of {\it weakly} isolated horizons (WIHs) as quasilocal 
	generalizations of event horizons is purely geometrical, and is independent of the variables used 
	in describing the gravitational field. 
	Here we consider a canonical decomposition of general relativity in terms of connection and vierbein variables starting from a first order action. Within this approach, the information about the existence of a (weakly) isolated horizon is obtained through a set of boundary conditions on an internal boundary of the spacetime region under consideration. We employ, for the self-dual action,  a generalization of the Dirac algorithm for regions with boundary. While the formalism for treating gauge theories with boundaries is unambiguous, the choice of dynamical variables on the boundary is not.
	We explore this freedom and consider different canonical formulations for non-rotating black holes as defined by WIHs. We show that both the notion of horizon degrees of freedom and energy associated to the horizon is not unique, even when the descriptions might be self-consistent. This represents a generalization of previous work on isolated horizons both in the exploration of this freedom and in the type of horizons considered. We comment on previous results found in the literature.
\end{abstract}


\maketitle
\tableofcontents

\section{Introduction}
Gauge theories defined in regions with boundaries have been the subject of extensive study from many different perspectives. 
To mention just a few of them: The boundary degrees of freedom  and dynamics and symmetries on the boundary, 
 conserved charges related to the boundary, consistency conditions on the boundary that the theory has to satisfy in order to be well defined  (see, for example \cite{RT,Banados1996,Troessaert,HM,BCE,ABOL} and references therein), a holographic principle that allows one to relate  a bulk theory with a boundary one \cite{Maldacena,HSS,MannMarolf}, and many more \cite{Barbero,Barbero1,Zabzine,MPB,Irais}. 
It is no overstatement to affirm that the proper understanding of boundary contributions to a physical theory is of the uttermost importance. The purpose of this manuscript is twofold. On the one hand, we are interested in exploring physical systems that can be analyzed from the perspective of a newly revised extension of the Dirac algorithm for theories defined with a region with a boundary \cite{CV-M+P}. Furthermore, as an application of the formalism, we are interested in making contact with research  on isolated horizons and the corresponding Hamiltonian description on them. As we shall now elaborate, these two issues become intertwined.
 
The particular gauge theory we shall consider is general relativity in a first order formulation. Within general relativity,
the concept of an isolated horizon was introduced in order to give a quasi-local description of a black hole in equilibrium \cite{PRL,ABL,ABLgeometry}. It turns out to be very useful to approach the quantum theory that finally leads to the calculation of the entropy of a black hole \cite{ABCKprl,ENPprl,ABKquantum}. From the classical point of view, we have a theory defined by an action in terms of tetrads or soldering forms and connections and want to account for the presence of the horizon that is modeled as an internal boundary to the spacetime region. 
The starting point in this case is a covariant action principle. The information about the weakly isolated horizon (WIH) on the boundary is included in the boundary conditions on the fields in the action principle.
In order to arrive at a quantum description, one needs to procure  a Hamiltonian formalism. However, the choice is not unique.
For example, it is well known that there are (at least) two Hamiltonian formalisms we can use: the better known canonical formulation, or the covariant one. 
In the first one the phase space is defined after a 3+1 decomposition of the theory and the corresponding Legendre transformation. In the second one the covariant phase space is defined as the space of solutions to the equations of motion; there is no splitting of spacetime.  

Both formulations have been applied to the issue at hand. Historically, the canonical approach was used in the early works on this subject \cite{ACKclassical,ABF0,ABF}, and also in  \cite{Thiemann:1993zq,ThiemannBook,CorichiReyes} for the asymptotically flat case, while the covariant one has received increased attention starting with \cite{AFK} and in most papers after that, see, for example \cite{AshtekarES,EngleNouiPerezPranzetti,ChatterjeeGhosh,crv1,CRGV2016,CRVwih2}.
 In both cases it was possible to obtain a consistent description that allowed for the computation of expressions for the mass of a black hole and to derive the zeroth and  first laws of the black hole mechanics.   Also, in the early work, a black hole was modeled using an isolated horizon, whose conditions have been since relaxed to the less restrictive weakly isolated horizon.     

Here, we will adopt the canonical approach and analyze a first order gravity action in self-dual variables in the presence of a non-rotating WIH. Our interest in this problem is twofold: 
First, we want to revisit and extend the results of \cite{ACKclassical} and \cite{ABF}, that were obtained for the case of spherically symmetric strongly isolated horizon, with a preferred foliation. In \cite{ACKclassical}, the phase space was also restricted to  configurations with the same horizon area, that led to a Chern-Simon action and the corresponding structure on the boundary. In \cite{ABF}, the area of the horizon was not fixed, which allowed for the derivation of the first law of black hole mechanics. Here, we want to further relax the conditions of \cite{ABF}, so that our phase space includes configurations with non-rotating WIH as an internal boundary. 
 
Furthermore, this manuscript can be seen as a continuation of the work on gauge theories defined in regions with boundaries, following the ideas of \cite{CV-P, CV-M+P}. There it was shown that when the action contains a boundary contribution with a time derivative, then a boundary contribution to the symplectic structure arises rather naturally. This also leads to a modification of the fundamental equations that now may contain contributions from the boundary.
In contrast to the examples studied in those works, here  the starting  covariant action does not have any boundary terms that contain time derivative of the fields, but they appear in the corresponding canonical action. The main result in this manuscript is that one may find several, different, descriptions of the canonical action and the corresponding degrees of freedom at the boundary.
As we will show in detail,  due to WIH boundary conditions, these boundary terms can be interpreted in different ways, for instance as a part of the boundary symplectic structure, or as a boundary term in the Hamiltonian.  Depending on how one chooses to interpret the terms one has a different canonical description of the system.

 The paper is organized as follows.  In section II we present some preliminary material. We start by recalling the basic ideas of the canonical treatment  of gauge theories defined in regions with boundary. Then, we briefly  review  the WIH boundary conditions and recall the spinorial formalism, in order to introduce our basic self-dual variables. 
 
In section III we derive the canonical action in self-dual variables, in the presence of boundaries, starting from the standard covariant self-dual action. We obtain the explicit form of a boundary term, for an arbitrary null boundary.  We also reproduce the expressions for the Hamiltonian, vector and Gauss constraints.
 
 In section IV we first study the compatibility with 3+1 foliations and the so called `comoving gauge', introduced in \cite{CRVwih2}, that will allow us to explicitly write down the expressions for boundary terms on WIH. 
 We show that it is only the Gauss constraint that contributes to the boundary equations of motion.
 
 In section V we present our main results. As mentioned earlier, it turns out that the boundary term that appears in the canonical action, can be interpreted as a contribution to the kinetic term, and thus to the boundary symplectic structure, or as a boundary term in the corresponding Hamiltonian. These options correspond to different choices of the boundary phase space. Once this election is made, we can obtain a well defined Hamiltonian, that is unique. We will analyze each of these settings in detail. 
 We will also show that the boundary term in the Hamiltonian cannot always be interpreted as the energy (mass) of the horizon.
 In these different situations, we also find different gauge symmetries on the horizon. In some of the cases there is a residual $U(1)$ gauge symmetry, 
 generated by the Gauss constraint that is defined in the bulk (where it is a generator of $SU(2)$ gauge symmetry), but whose functional derivative has contribution on the horizon. We end by commenting on the different choices that one has to make to have consistent descriptions of the system.
 
In Appendix A we look into some additional details of the potentials that in some cases represent boundary degrees of freedom. In Appendix B we reproduce the results obtained earlier in the spherically symmetric case of strongly isolated non-rotating horizon and the phase space with fixed horizon area \cite{ACKclassical,ABF}. We obtain the Chern-Simons boundary symplectic structure and show that in this case, tangential diffeomorfisms {\it are}  boundary gauge symmetries of the theory.

Throughout the manuscript, we assume the spacetime to be a four dimensional smooth manifold $\mathcal{M}$.  Furthermore, we will take $\mathcal{M}$, or at least a portion of it $M\subseteq\mathcal{M}$, to be globally hyperbolic, with a boundary $\partial M$,  and such that it may be foliated as $M\approx\mathbb{R}\times\Sigma$, with $\Sigma$ a spatial hypersurface with inner boundary homeomorphic to a 2-sphere. We will use greek letters to denote abstract (or actual component) spacetime indices running from 0 to 3. Similarly, lowercase latin letters from the beginning of the alphabet will correspond to abstract or component spatial indices on $\Sigma$ running from 1 to 3.

\section{Preliminaries}

In this section we shall present some preliminary material in order to make the manuscript self contained. It has three parts. In the first one, we briefly summarize the formalism of gauge theories in the presence of a boundary as developed in \cite{CV-M+P}. In the second one, we summarize the isolated horizon boundary conditions and its main geometrical constituents. In the last part we introduce the basics of the spinorial formalism needed to write down the self-dual action that we shall consider in the rest of the manuscript. Readers familiar with either of these topics may skip the corresponding part of this section.

\subsection{Canonical formalism for gauge theories defined in regions with boundary}

 We shall start by recalling some of the main results of \cite{CV-M+P}. 
In that contribution the objective was to study gauge theories that could have a boundary term in the covariant (or canonical) action, containing time derivatives of the fields. It was shown that the symplectic structure can acquire a boundary contribution. In that case the standard Dirac algorithm has to be extended in order to have a consistent description, that includes an extended phase space with a boundary phase space, degrees of freedom and (possibly) constraints on the boundary. 
 
 
In this approach the starting point is a canonical action that is obtained after a 3+1 decomposition of a covariant action. 

To be more concrete, let us suppose that as a result of the 3+1 decomposition, the canonical action has the form
 \be
S[A,P]=\int \left\{ P[{\cal{L}}_t{A}] - H\right\} \md t\, ,
\ee
where $P[{\cal{L}}_t{A}]$ is the kinetic term or symplectic potential\footnote{More precisely, one uses $P[{\cal{L}}_t{A}]$ (which is a one-form on configuration space) to define or identify the symplectic potential and symplectic structure on phase space. For details see \cite{CV-M+P}. 
}, whose general form in the space region with a boundary is
\be
P[{\cal{L}}_t{A}] = \int_\Sigma\md^3x\,\,  \tilde{P}^a_{\;BC}\, {\cal{L}}_t A_a^{\;BC} + 
\int_{\partial\Sigma}\md^2y\,\, \tilde{\pi}^j\, {\cal{L}}_t \alpha_j\, .
\ee
$A_a^{\;BC}(x)$ are the bulk field configuration variables, and $\tilde{P}^a_{\;BC}(x)$ their corresponding momenta,  $a$ is a spatial index and $(B,C)$ are internal indices.\footnote{Customary, one denotes internal indices in gauge theories in terms of $I$, an index for the Lie algebra of the gauge group. Here, for convenience and without loss of generality, we shall use a pair of spinorial indices $AB$, since that shall be the structure of the internal space we will consider.}
$\alpha_j(y)$ are to be interpreted as boundary configuration variables, and $ \tilde{\pi}^j(y)$ as their corresponding momenta\footnote{We are denoting by $j$ all possible index structure that the fields at the boundary have.}.
In the general case, $H$ is a candidate for a Hamiltonian that may depend also on the boundary variables $(\alpha_j, \tilde{\pi}^j)$, on top of the bulk canonical variables restricted to the boundary. In this manuscript we will show that the boundary term in $P[{\cal{L}}_t{A}]$, due to given boundary conditions, is not unique.
It can be  chosen in different ways, leading  to a different  boundary phase space in each case. In every one of those cases,  we shall find a corresponding well defined Hamiltonian. 
 
The kinetic term
determines the symplectic structure of the theory as \cite{CV-M+P} :
 \be
\Omega= \int_\Sigma\md^3x\,\,\ed\tilde{P}^a_{\;BC}\dwedge\; \ed A_a^{\;BC}  + \int_{\partial\Sigma}\md^2x\,\, \ed\tilde{\pi}^j\dwedge\; \ed \alpha_j\, ,\label{omega-with-boundary}
\ee
with $\ed$ and {$\wedge \!\!\!\!\! \wedge\,$} the differential and wedge operator respectively on (infinite dimensional) phase space.
We are regarding the bulk and boundary degrees of freedom as independent, even when one (or more) of the boundary fields might have come from some appropriate restriction of bulk degrees of freedom, that is imposed by boundary conditions.

The function $H$ defines a Hamiltonian vector field (HVF) $X_H$ through the equation 
 \be\label{dH}
 \ed H(Y) = \Omega (Y, X_H)\, .
 \ee
 Given the boundary contribution to the symplectic structure $\Omega$, the bulk part of (\ref{dH}) gives the equations of motion in the bulk, while
the boundary term  defines the boundary components of the Hamiltonian vector field (HVF).
 We will call a corresponding canonical description consistent if: a) in the case when there are no boundary contributions to the symplectic structure, the boundary term in $\ed H$ vanishes or b) in the case when there are boundary contributions to the symplectic structure, $X_H$ generates consistent evolution of both the bulk and boundary degrees of freedom.

In gauge theories we have a certain set of (smeared) constraints $G_I[f]$ and in that case there is an additional set of consistency conditions:
\be\label{tangX}
\ed G_I[f](X_H)\approx 0\, .
\ee
These conditions imply that $X_H$ is tangent to the constraint surface.

To write down explicitly the expressions (\ref{dH}) and (\ref{tangX}) we need to write down the  generic HVF \cite{CV-M+P}
\begin{eqnarray}
X &=& \int_\Sigma\md^3x\,\left[ (X_A)_a^{\;BC}(x)\left(\frac{\delta}{\delta A_a^{\;BC}(x)}\right)+
(X_{\tilde{P}})^a_{\;BC}(x)\left(\frac{\delta}{\delta {\tilde{P}}^a_{\;BC}(x)}\right)\right] \nonumber\\
&+& \int_{\partial\Sigma}
\md^2 y\,\left[ (X_{\alpha})_j(y)\left(\frac{\delta}{\delta\alpha_j(y)}\right)+
(X_{\tilde{\pi}})^j (y)\left(\frac{\delta}{\delta {\tilde{\pi}}^j(y)}\right)\right]
\, ,\label{genericvectorboundary}
\end{eqnarray}
whose components in the bulk are $(X_{A}, X_{\tilde{P}})$ and on the boundary $(X_{\alpha}, X_{\tilde{\pi}})$.

The gradient of an eligible function $F$ will have the form,
\begin{eqnarray}
\ed F &=& \int_\Sigma \md^3\!x\;\left(\frac{\delta F}{\delta A_a^{\;BC} (x)}\,\ed A_a^{\;BC}(x)+
\frac{\delta F}{\delta {\tilde{P}}^a_{\;BC} (x)}\,\ed {\tilde{P}}^a_{\;BC}(x)\right)\nonumber \\
&{}& + \int_{\partial\Sigma}\md^2\!y\;\left(\frac{\delta F}{\delta \alpha_j(y)}\,\ed \alpha_j(y)
+\frac{\delta F}{\delta {\tilde{\pi}}^j(y)}\,\ed {\tilde{\pi}}^j(y)\right)\, , \label{gradconfrontera}
\end{eqnarray}
with its corresponding contribution from the boundary. 
As usual, to find these gradients we will identify $\ed F(Y)$ with variations of $F$ along the direction of $Y$, i.e. $\ed F(Y)=\delta_Y F$.
The quantities $\frac{\delta F}{\delta \alpha_j(y)}$ should be thought as a generalization of the partial derivative of the function $F$ along the coordinate $\alpha_j$ and leaving the rest of the coordinates (including the bulk DOF) constant. It is important to note that, since the function $F$ depends on all possible variables (bulk and boundary), the boundary contribution to the gradient might depend explicitly on the bulk variables, restricted to the boundary, as well. 

From (\ref{omega-with-boundary}) and (\ref{genericvectorboundary}) it follows that
\begin{equation}
 \Omega (Y,X) = \int_\Sigma\md^3x\, ( X_A Y_{\tilde{P}} -
 Y_A X_{\tilde{P}} )
 + \int_{\partial\Sigma}\md^2 y\, (X_{\alpha}Y_{\tilde{\pi}}-Y_{\alpha}X_{\tilde{\pi}})\, ,\label{BoundarySS}
\end{equation}
where the contraction over the corresponding indexes is understood. 

Here we are interested  in the theory of gravity in the first order formalism, that is a diffeomorphism invariant theory. Its bulk Hamiltonian is a linear combination of the Hamiltonian, vector and Gauss constraints, so the standard interpretation is that the bulk part of the Hamiltonian only generates gauge transformations (in the Hamiltonian definition of gauge, as explained in \cite{CV-M+P}). As is well known, in the case of asymptotically flat spacetimes, in order to obtain a well defined Hamiltonian, we need to add an asymptotic boundary term  whose numerical evaluation corresponds to the ADM mass \cite{RT,CorichiReyes}. This functional is no longer a constraint but a true generator of asymptotic time translations.

There is an important remark to be made about the use of  the terms {\it gauge} and {\it gauge symmetry} from the viewpoint of Hamiltonian dynamics. We are considering a framework where the fundamental variables are to be taken as connections with their corresponding conjugate momenta, where the Lie algebra where they live corresponds to the Lorentz 
Lie algebra. Thus, there is a natural notion of gauge in the sense of Lorentz transformations, and invariance. There is also the Hamiltonian notion of gauge, where gauge transformation are to be generated only by first class constraints. Sometimes in the text that follows we shall simply refer to {\it gauge}, and we hope that the context will be enough to differentiate between the two notions.

 As a natural assumption, in \cite{ABF}, the mass of an isolated horizon was defined as the numerical, on-shell value of the horizon boundary term in the corresponding Hamiltonian. The result was identical to the Smarr expression for the mass of a non-rotating black hole \cite{Smarr}. A priori, it is not obvious that it would be the case, since the horizon is a null surface, in contrast to a time-like asymptotic region where the boundary is time-like. We will show that, depending on the choice of a boundary phase space, we can obtain a consistent description of WIH dynamics, even in the case when the Hamiltonian does not have a horizon boundary term. We will also show under which assumptions we can obtain the expected expression for the mass of a black hole. 
 
We should note that, in a somewhat independent manner, there have been several approaches to gravitational degrees of freedom on a null surface. Several examples
 can be found in \cite{Wieland1,HopfmuellerFreidel,padmadhavan,Flanagan}. Our approach is different from those.

\subsection{WIH boundary conditions} 

In this part we shall give definitions and some of the basic properties of isolated horizons.
For more details see for instance, \cite{AFK} and \cite{CRVwih2}.
Isolated horizons are modeled by special types of null hypersurfaces on $\mathcal{M}$.  A three dimensional hypersurface $\Delta\subset\mathcal{M}$ is \emph{null} if its induced metric $h_\pbi{\mu\nu}$ is degenerate  of signature $(0,+,+)$ (following previous conventions, we will denote forms
pulled back to $\Delta$ using indexes with arrows under them).
This degeneracy is equivalent to the requirement that any vector $\ell^\mu$ normal to $\Delta$ is null: $\ell_\mu\ell^\mu=0$, and it follows that a fundamental property of null hypersurfaces is that they are ruled by null geodesics. Any null normal vector field $\ell^\mu$ on $\Delta$ satisfies the geodesic equation
\[
\ell^\mu\nabla_\mu\ell^\nu=\kappa_{(\ell)}\ell^\nu\,,
\]
and the non-affinity parameter $\kappa_{(\ell)}$ will 
correspond to surface gravity when one specializes to isolated horizons. 

For null hypersurfaces, it is convenient to use the null normal $\ell^\mu$  at each point $p\in\Delta$ to construct or fix a \emph{Newman-Penrose null basis}  $(k,\ell,m,\bar{m})$ on $T_p\mathcal{M}$ \cite{NewmanPenrose}.
 This is done by choosing a  null direction $k^\mu$ transverse to $T_p\Delta$ on the light cone at each $T_p\mathcal{M}$, normalized such that $k_\mu\ell^\mu=-1$. The orthogonal complement 
 of the plane spanned by $k^\mu$ and $\ell^\mu$ is a two dimensional spatial subspace of $T_p\Delta$, 
 where one can choose a  null basis given by a complex vector $m^\mu$ and its conjugate $\bar{m}$, such that $m_\mu\bar{m}^\mu=1$. 
We will call a basis constructed in this way a null basis \emph{adapted to} $\Delta$.

A null basis adapted to $\Delta$ is not unique,
but any two adapted null bases are related by 
a Lorentz rotation preserving the direction of $\ell^\mu$. 
The \emph{cross sectional area two-form} 
$\leftidx{^2}{\epsilon}{}:=im\wedge\bar{m}\,$,
is invariantly defined on $\Delta$, in the sense that it is the same regardless of the choice of adapted null basis.
Similarly, the (two dimensional spatial) \emph{cross sectional metric}  
$2\,m_{(\pbi{\mu}}\bar{m}_{\pbi{\nu})}=h_\pbi{\mu\nu}\,$
is invariantly defined on $T_p\Delta$ and equal to the pullback metric.

The null vector $k^\mu$ allows one to define a projector to $T_p\Delta$, 
$
\Pi^\mu_{\;\;\nu}:=\delta^\mu_{\;\;\nu}+k^\mu\ell_\nu
$.
The second fundamental form of $\Delta$ with respect to $\ell^\mu$ 
is defined as
\begin{equation} \label{ThetaDef}
\Theta_{\mu\nu}:=\Pi^\sigma_{\;\;\mu}\Pi^\rho_{\;\;\nu}\nabla_\sigma\ell_\rho =\nabla_\mu\ell_\nu-\omega_\mu\ell_\nu+\ell_\mu k^\sigma\nabla_\sigma\ell_\nu\,,
\end{equation}
with
\begin{equation}  \label{omegaDef}
\omega_\mu:=-k^\sigma\nabla_\mu\ell_\sigma-\ell_\mu k_\sigma k^\rho\nabla_\rho\ell^\sigma\,,
\end{equation}
defined as the \emph{rotation 1-form} in the context of black hole horizons. Contracting this last definition with $\ell^\mu$, it follows that 
\be
\kappa_{(\ell)}=\omega_\mu\ell^\mu.
\ee 
The second fundamental form $\Theta_{\mu\nu}$ encodes the `kinematics' of the geodesic congruence of null generators of $\Delta$ with velocities $\ell^\mu$. Its trace $\theta_{(\ell)}:=g^{\mu\nu}\Theta_{\mu\nu}$ defines the \emph{expansion} of the congruence and its symmetric trace-free part is the \emph{shear}. The \emph{twist} or anti-symmetric part vanishes as a consequence of the congruence being hypersurface orthogonal.

To model equilibrium horizons resulting from gravitational collapse, one first restricts the topology of $\Delta$ and incorporates the notion that the null geodesic generators should be non-expanding.

A null hypersurface $\Delta\subset\mathcal{M}$ is called a \emph{non-expanding horizon} (NEH) if
\begin{enumerate}[(i)]
\item $\Delta$ is diffeomorphic to the product $S^2\times\mathbb{R}$ with the fibers of the canonical projection $P:S^2\times\mathbb{R}\to S^2$ corresponding to the null generators.
\item The expansion $\theta_{(\ell)}$ of any null normal $\ell^\mu$ to $\Delta$ vanishes.
\item On $\Delta$, Einstein's equations hold and the stress-energy tensor $T_{\mu\nu}$ of matter satisfies the \emph{null dominant energy condition}, i.e. $-T^\mu_{\;\;\nu}\,\ell^\nu$ is causal and future-directed.
\end{enumerate}

These conditions imply  that  on $\Delta$ 
the complete second fundamental form $\Theta_{\mu\nu}$ vanishes \cite{AFK,GJreview}. So from (\ref{ThetaDef}) and $\ell_\pbi{\mu}=0$, it follows that on a NEH
\begin{equation} \label{rotationForm}
\nabla_{\pbi{\mu}}\ell^\nu\WIHeq\omega_\pbi{\mu}\ell^\nu\,,
\end{equation}
where we used the standard notation $\WIHeq$ for equalities valid only on $\Delta$.

Equation (\ref{rotationForm}) also implies that the induced metric on $\Delta$ and the transverse area 2-form are Lie dragged along $\ell^\mu$ ($\ell^\mu$ is a symmetry direction for the metric so NEHs are a generalization of Killing horizons, and the horizon area is preserved):
\begin{equation} \label{Lh}
\mathcal{L}_{\ell}\,h_{\pbi{\mu\nu}}\WIHeq 0\,  \qquad\text{ and }\qquad  \mathcal{L}_{\ell}\,\leftidx{^2}{\epsilon}{}\WIHeq 0\,.
\end{equation}
In a general null hypersurface $\Delta$ each choice of $k^\mu$ defines 
a (torsion-free) induced connection $\widehat{\nabla}_\pbi{\mu}$ compatible with the induced metric: $\widehat{\nabla}_\pbi{\mu}h_{\pbi{\nu\rho}}=0$. On a NEH this connection is unique and `intrinsic' to $\Delta$ (independent of the choice of $k^\mu$),
but unlike spatial or time-like surfaces, on a NEH the connection is not fully determined by the metric. It is hence the pair $\big(h_{\pbi{\mu\nu}},\widehat{\nabla}_\pbi{\mu}\big)$ taken together that defines the \emph{intrinsic geometry} on $\Delta$.

Along with  $\omega_\pbi{\mu}$, on a NEH there is another relevant intrinsic one-form arising from the connection on $\Delta$, the \emph{transverse connection potential} $V_\pbi{\mu}$. 
Using an adapted null basis, it is  defined as:
\begin{equation} \label{tV}
V_\pbi{\mu}:=\bar{m}_\nu\widehat{\nabla}_\pbi{\mu}m^\nu\,.
\end{equation}
For restricted Lorentz transformations of the null basis, 
this expression is invariantly defined except for $U(1)$-rotations for which it transforms as a $U(1)$-connection.

The definition of NEHs is too general to capture the mechanics of black holes.
To derive the zeroth and first laws, one needs to further select an equivalence class of normal vector fields on $\Delta$.

A \emph{weakly isolated horizon} $(\Delta,[\ell])$ is a non-expanding horizon where an equivalence class of null normals $[\ell]$ satisfying  
\begin{equation} \label{WIHcondition}
\mathcal{L}_{{\ell}}\,\omega_\pbi{\mu}\WIHeq 0  \qquad \text{ for all } \; \ell^\mu\in[\ell]
\end{equation}
has been singled out. Two normals $\tilde{\ell}\sim\ell$ belong to the same equivalence class iff $\tilde{\ell}^\mu=c\ell^\mu$ for some constant $c>0$ on $\Delta$.  

This symmetry condition or time-independence of a part of the induced connection
 is sufficient to ensure that the surface gravity $\kappa_{(\ell)}$ is constant on a WIH, that constitutes the zeroth law of black hole mechanics.

Every NEH can be made into a WIH by  appropriate (albeit infinitely many different) choices of equivalence classes $[\ell]$.  We will hence work mainly within the class of weakly isolated horizons in this paper.

\subsection{Spinorial formulation}

Our goal is to derive the canonical Hamiltonian formulation for general WIHs in self-dual variables starting from the covariant self-dual action. Therefore, 
in this section we present  a brief review of the spinorial formalism, in order to make this manuscript more self-contained. Details can be consulted in \cite{ABF0, ABF, ACKclassical, CRVwih2}. We follow the same conventions used in \cite{ThiemannBook, CRVwih2}.

For the spinorial self-dual formulation of gravity the basic variables are a pair $(\sigma_\mu^{AA'},\leftidx{^+}{\A}{_{AB}})$, where $\sigma_\mu^{AA'}$ is a soldering form for primed and unprimed $SL(2,\mathbb{C})$ spinors and $\leftidx{^+}{\A}{_{AB}}$ is a $SL(2,\mathbb{C})$ self-dual connection.

The spinor bundle is the vector bundle over $\mathcal{M}$ whose fiber is the complex two dimensional vector space $\mathbb{C}^2$, with
symplectic structure or metric $\epsilon_{AB}$ with inverse $\epsilon^{AB}$, such that  $\epsilon_{AB}=-\epsilon_{BA}$ and  $\epsilon^{AB}\epsilon_{AC}=\delta^B_C$.
($A$, $B$, $C$ are spinor indices running from 1 to 2.) 
$\epsilon_{AB}$ and $\epsilon^{AB}$ are used
 to raise and lower spinor indices:  $X_B:=X^A\epsilon_{AB}$ and $X^A=\epsilon^{AB}X_B$, which implies $X^AY_A=-X_AY^A$.  
Complex conjugation $X^A\mapsto\bar{X}^{A'}$, maps spinors to the \emph{conjugate} spinor space with metric $\bar{\epsilon}_{A'B'}$.

The \emph{soldering vector} $\sigma^\mu_{AA'}$ provides an isomorphism between the space of anti-hermitian spinors $X^{AA'}$ (such that $\bar{X}^{AA'}=-X^{A'A}$) and the tangent space $T_p\mathcal{M}$ at each point $p$. Its inverse, the \emph{soldering form} $\sigma_\mu^{AB}$, is obtained by lowering spacetime indices with $g_{\mu\nu}$ and raising spinor indices with $\epsilon^{AB}$ and $\bar{\epsilon}^{A'B'}$.
They satisfy
\be
\sigma_\mu^{AA'}\sigma^\nu_{AA'}=\delta^\nu_\mu,  \qquad  \sigma_\mu^{AA'}\sigma^\mu_{BB'}=\delta^{A}_{B}\delta^{A'}_{B'}
\qquad \text{ and }  \qquad    \bar{\sigma}^\mu_{AA'}=-\sigma^\mu_{A'A}\,.
\ee

For convenience, one can introduce a fixed dyad basis $(\om^A,\iota^A)$ in spinor space $\mathbb{C}^2$ at each point, such that  
\be
\iota^A\iota_A=0\, ,   \qquad\qquad  \om^A\om_A =0\, ,  \qquad\qquad \iota^A\om_A=-\iota_A\om^A=1\, ,
\ee
and similarly for the conjugate basis $(\bar{\om}^{A'},\bar{\iota}^{A'})$. We assume our auxiliary connection is such that $\bar\partial(\iota^A,\om^A)=0$, and we will fix this internal dyad  such that $(\delta\iota^A,\delta\om^A)=0$.

The spinors $\{i\,\om^A\bar{\om}^{A'},i\,\iota^A\bar{\iota}^{A'},i\,\om^A\bar{\iota}^{A'},i\,\iota^A\bar{\om}^{A'}\}$ define a basis for the internal space of anti-hermitian spinors $X^{AA'}$.
One can define a  Newman-Penrose null basis on the tangent space:
\begin{align}
l^\mu:=\sigma^\mu_{AA'}\,l^{AA'}:=i\sigma^\mu_{AA'}\om^A\bar{\om}^{A'} \, ,   \qquad \qquad 
 &m^\mu:=\sigma^\mu_{AA'}\,m^{AA'}:=i\sigma^\mu_{AA'}\om^A\bar{\iota}^{A'} \, , \notag\\
k^\mu:=\sigma^\mu_{AA'}\,k^{AA'}:=i\sigma^\mu_{AA'}\iota^A\bar{\iota}^{A'} \, , \qquad \qquad  
&\bar{m}^\mu:=\sigma^\mu_{AA'}\,\bar{m}^{AA'}:=i\sigma^\mu_{AA'}\iota^A\bar{\om}^{A'}\, , \label{solderingIsomorphism}
\end{align}
The soldering form is hence expanded as
\begin{equation}  \label{SL2CsolderingExpansion}
\sigma^\mu_{AA'}=-k^\mu l_{AA'}- l^\mu k_{AA'}+\bar{m}^\mu m_{AA'}+m^\mu \bar{m}_{AA'}\, .
\end{equation}

On $\Delta$ one can define 
the spin connection compatible with the soldering form or vector, $\nabla_\mu\sigma^\nu_{AA'}=0$, such that
\[
\nabla_\mu X^{AA'}=\bar{\partial}_\mu X^{AA'}+\leftidx{^+}{\A}{^A_{\;\;B}}X^{BA'}+\leftidx{^-}{\bar{\A}}{^{A'}_{\;\;B'}}X^{AB'}\, ,
\]
where $\leftidx{^+}{\A}{^A_{\;\;B}}$ is the self-dual connection potential and  
$\leftidx{^-}{\bar{\A}}{^{A'}_{\;\;B'}}$ is the conjugate of the anti self-dual connection potential, while $\bar{\partial}$ is a fiducial flat connection.
Expanding  covariant derivatives of the null tetrad in terms of the spinor basis and using compatibility condition, one can write down an expansion for the self-dual connection potential in terms of the one-forms $(\W ,\V ,\U ,\Y)$ and the dyad $(\iota_A,\om_A)$:
\begin{equation}\label{SDAexpansion}
\leftidx{^+}{\A}{_{AB}}=-(\W+\V)\,\iota_{(A}\om_{B)}+\bar{\U}\,\om_A\om_B-\Y\,\iota_A\iota_B\, ,
\end{equation}
where
\begin{equation} \label{WVUYdef}
\W_\mu:=-k_\nu\nabla_\mu l^\nu\,, \qquad \V_\mu:=\bar{m}_\nu\nabla_\mu m^\nu\,, \qquad \U_\mu:=m_\nu\nabla_\mu k^\nu\,, \qquad \Y_\mu:=m_\nu\nabla_\mu l^\nu\,.
\end{equation}

Expansion (\ref{SL2CsolderingExpansion}) is valid on the bulk as well as on $\Delta$, while the expansion (\ref{SDAexpansion}) is valid only on $\Delta$. Both are independent of any gauge choice.


\section{Canonical action in self-dual variables}
In this section we shall review the canonical decomposition of the standard first order self-dual action to arrive at the canonical action, following in detail all the boundary contributions.

Spherically symmetric isolated horizons were originally defined in the spinorial formalism, see, for example \cite{ABF, ACKclassical}, which turned out to be especially convenient. The canonical analysis of first order gravity theories  in self-dual spinorial variables was first performed in \cite{Samuel,JacobsonSmolin}. The classical phase space in theories with the spherically (strongly) isolated horizon, with fixed area, was studied for the first time in \cite{ACKclassical}.
The boundary symplectic structure in this case coincides with the one of the $U(1)$ Chern-Simons theory (such theory was then quantized in \cite{ABCKprl,ABKquantum}). In \cite{ABF} this analysis was extended to include the phase space where the area of the horizon could change, leading to the generalization of the equilibrium version of the first law of black hole mechanics. The corresponding boundary symplectic structure in this case does {\it not} correspond to a Chern-Simons theory.

We start from the same covariant action, and perform its 3+1 decomposition, in a more general case, where the internal boundary is a weakly isolated horizon. We will show that, 
 there is not a unique election of the boundary variables. In what follows, we will explore the consequences of these different choices. One of them extends the results of \cite{ABF}, but the other ones also lead to consistent canonical descriptions.

To derive a canonical action and a Hamiltonian theory consistent with WIH boundary conditions, we start with the first order covariant action for gravity in self-dual variables, given by (see, for example, \cite{Samuel,JacobsonSmolin,AshtekarLectures,ACKclassical,ABF})
 \begin{align}
 S_{\text{SD}}(\sigma,\sdA):=\int_{M}\md^4x \,(-\,\leftidx{^4}{\sigma})\sigma^\mu_{\;A}\,^{A'}\sigma^\nu_{\;BA'}\sdF_{\mu\nu}^{\,AB}  \, ,     \label{SDactionComponentForm}
  \end{align}
where $\leftidx{^4}{\sigma}=\det{(\sigma^\mu_{\;A}\,^{A'})}=\sqrt{-g}$, with $g=\det{(g_{\mu\nu})}$ and 
\be
\sdF_{\mu\nu}^{\,AB}=2\,\partial_{[\mu}\sdA_{\nu ]}^{\;AB}+\sdA_{[\mu\ C}^{\ A}\sdA_{\nu ]}^{\,CB}\, ,
\ee
is the curvature of the self-dual connection. Without loss of generality we are fixing the value of the coupling constant to 1. 
For asymtotically flat spacetimes,
we are neglecting the boundary term in the asymptotic region, that should be added in order to make the action differentiable at infinity \cite{ABF, Thiemann:1993zq, CorichiReyes}. Here we are interested in the inner boundary, WIH, and in what follows we will not take into account the contributions from the asymptotic region, as they are well understood.
For asymptotically flat spacetimes with a WIH as inner boundary, the self-dual action (\ref{SDactionComponentForm}) (with boundary term at infinity) is well defined, i.e. finite, and differentiable in the sense that its variations vanish at the asymptotic and inner boundaries. This is true for arbitrary variations with the full $SL(2,\mathbb{C})$ gauge freedom at the WIH\footnote{Due to the presence of terms of the form $\ell\cdot \delta V$ at the horizon, in \cite{CRVwih2}, it was stated that differentiability of the action required a partial gauge fixing and/or a counter term at $\Delta$. There is however a subtle argument to  show that these terms actually vanish.}.

For a 3+1 decomposition and Hamiltonian formulation of a covariant theory, one postulates a time function $t:M\subseteq\mathcal{M}\to\mathbb{R}$ on (a portion of) spacetime $\mathcal{M}$, whose level curves $\Sigma_t$ are spatial hypersurfaces and provide a foliation of the spacetime region $M$. Spacetime fields are split into tangential (spatial) and normal components with respect to this foliation.  For the configuration space consisting of spacetimes admitting a weakly isolated horizon $\Delta$ as an internal boundary, we will only consider foliations of the bulk of $M$ whose intersection with $\Delta$ is non trivial and induces a foliation of $\Delta$ by two-spheres $\mathcal{S}_t=\Sigma_t\cap\Delta$.

Additionally, one needs to choose an \emph{evolution} vector field $t^\mu$ such that $t^\mu\nabla_\mu t=1$ and along which spatial fields are defined to `evolve'.  The evolution vector field will generically be time-like on the bulk. For non-rotating horizons it will be chosen to belong to the equivalence class $[\ell]$ on $\Delta$. That is, the vector $t^\mu$ becomes null as one approaches  $\Delta$.

Decomposition of the spacetime metric splits the ten independent components of $g_{\mu\nu}$ into the six independent components of the Euclidean spatial metric $q_{\mu\nu}$, the lapse function $N$ and the  shift vector $N^\mu$. Lapse and shift being respectively the normal and tangential components of the evolution vector field $t^\mu=Nn^\mu+N^\mu$, where  $n^\mu$ denotes the future directed unit normal to the foliation and $N=-\left|(\md t)_\mu(\md t)^\mu\right|^{-1/2}$. In coordinates $(t,y^a)$ adapted to the foliation the line element reads
\begin{equation} \label{ADMmetric}
g_{\mu\nu}\md x^\mu \md x^\nu=(-N^2+q_{ab}N^aN^b)\md t^2+2q_{ab}N^b\md t\,\md y^a + q_{ab}\md y^a \,\md y^b\,.
\end{equation}

The four-dimensional  $SL(2,\mathbb{C})$ soldering form on $M$, $\sigma^\nu_{\;AA'}$, induces a three-dimensional   $SU(2)$ soldering form on $\Sigma$, $\sigma^\mu_{\;A}\,^B$, in the following way
\begin{equation} \label{SU2solderingForm}
\sigma^\mu_{\;A}\,^B:=-i\sqrt{2}\,q^\mu_{\,\nu}\,\sigma^\nu_{\;AA'}\,n^{A'B}\, ,
\end{equation}
where $q^\mu_{\,\nu}:= \delta^\mu_\nu + n^\mu n_\nu$ is a projector on $\Sigma$ and $n^{A'B}=n^\mu\sigma_\mu^{\;A'B}$ is the spinorial representation of $n^\mu$. From (\ref{SU2solderingForm}) it follows that
\begin{equation}
\sigma^\mu_{\;AA'}=-i\sqrt{2}\,\sigma^\mu_{\;A}\,^B\,n_{BA'}-n^\mu n_{AA'}.
\end{equation}

From this one gets the 3+1 decomposition (see, for example, \cite{Thiemann:1993zq})
\begin{align}
S_{\text{SD}}=\int\md t\,\md^3y\big\{& i\sqrt{2}\,\tilde{\sigma}^a_{\;AB}\,\mathcal{L}_{t}A_a^{\;AB}
-\frac{N}{\sqrt{q}}\,\tilde{\sigma}^a_{\;A}\,^C\,\tilde{\sigma}^b_{\;CB}\,F_{ab}^{\;AB}  \notag\\
&-i\sqrt{2}\,N^a\,\tilde{\sigma}^b_{\;AB}\,F_{ab}^{\;AB} 
+ (t^\mu\,\sdA_\mu^{\;AB})\,\leftidx{^A}{D}_a(i\sqrt{2}\,\tilde{\sigma}^a_{\;AB}) \big\}   \notag\\
-\int_M\md^4x\,\partial_\nu&\left[(t^\mu\,\sdA_\mu^{\;AB})\,(i\sqrt{2}\,\tilde{\sigma}^a_{\;AB}\,\tilde{e}^\nu_a)\right],
\label{SDaction3plus1}
\end{align}
with densitized soldering form
\[
\tilde{\sigma}^a_{\;AB}:=\sqrt{q}\;\sigma^a_{\;AB}=\sqrt{q}\;\tilde{e}^a_\mu\,\sigma^\mu_{\;AB},
\]
where $q=\det q_{ab}$, $\tilde{e}^\mu_a:=\Par{x^\mu}{y^a}$ are the components of tangent vectors to $\Sigma$ ($\tilde{e}^\mu_a n^a=0$) and $\tilde{e}^a_\mu = g_{\mu\nu}\, q^{ab}\,\tilde{e}^\nu_b$. Furthermore, 
\[
F_{ab}^{\;AB}:=\tilde{e}^\mu_a\,\tilde{e}^\nu_b\,\sdF_{\mu\nu}^{\;AB}=\Par{x^\mu}{y^a}\Par{x^\nu}{y^b}\,\sdF_{\mu\nu}^{\;AB},
\]
is the pull-back to $\Sigma$ of the self-dual curvature, which matches the curvature of the pulled back connection
\[
A_{a}^{\;AB}:=\tilde{e}^\mu_a\,\sdA_{\mu}^{\;AB}=\Par{x^\mu}{y^a}\,\sdA_{\mu}^{\;AB}\, ,
\]
and
\[
 \leftidx{^A}{D}_a \tilde{\sigma}^a_{\;AB}  = \partial_a\tilde{\sigma}^a_{\;AB}+{A_{aA}}^C\tilde{\sigma}^a_{\;CB}+{A_{aB}}^C\tilde{\sigma}^a_{\;AC}\, .
\]

From Stokes' theorem, the last term in the decomposition (\ref{SDaction3plus1}) gives a boundary term $S_{\text{SD}}^{\partial M}$. If we use the short hand notation $\tilde{B}^\nu:=-(t^\mu\,\sdA_\mu^{\;AB})\,(i\sqrt{2}\,\tilde{\sigma}^a_{\;AB}\,\tilde{e}^\nu_a)$ :
\[
\int_M\md^4x\,\partial_\nu\tilde{B}^\nu=\oint_{\partial M}\left(\frac{\tilde{B}^\nu}{\sqrt{-g}}\right)\md\mathcal{S}_\nu,
\]
with $\md\mathcal{S}_\nu$ the oriented volume form on $\partial M=\Delta\cup \Sigma_1\cup\Sigma_2\cup \tau_\infty$.

On the spatial hypersurfaces $\Sigma_{1,2}$, $\md\mathcal{S}_\nu=n_\nu\,\md^3y$, but since $\tilde{e}^\nu_a$ is tangent to $\Sigma$, the surface terms vanish there, so that the boundary term has only contributions from the WIH and from the asymptotic region. 
\[
S_{\text{SD}}^{\partial M}=S_{\text{SD}}^{\Delta}+S_{\text{SD}}^{\tau_\infty}\, .
\]
Here, we will neglect the last term, since its contribution can be seen in \cite{CorichiReyes}.
Focusing on the inner horizon boundary $\Delta$ and using adapted coordinates $z^{\pbi{\mu}}=(\lambda,\Theta_1,\Theta_2)$, with $\lambda$ parametrizing
 null geodesics $x^\mu(\lambda)$ such that null normals are $\ell^\mu=\frac{\md x^\mu}{\md\lambda}$,  its volume element is \cite{Poisson} 
\begin{equation} \label{DeltaVolume}
\md\mathcal{S}_\nu=-\ell_\nu\leftidx{^2}{\epsilon}\,\md\lambda \, ,
\end{equation}
where $\leftidx{^2}{\epsilon} = \sqrt{h}\,\md^2\Theta$ is the transverse area two form. Then, the WIH boundary term can be written as
\begin{equation} \label{SDactionBoundary}
S_{\text{SD}}^{\Delta}=\int_\Delta\,\frac{1}{N}\, (t^\mu\,\sdA_\mu^{\;AB})\,(i\sqrt{2}\,{\sigma}^a_{\;AB}\,\tilde{e}^\nu_a) \ell_\nu\leftidx{^2}{\epsilon}\,\md\lambda ,
\end{equation}
where we have used $\sqrt{-g}=N\sqrt{q}$.

Action (\ref{SDaction3plus1}) has the standard canonical form:
\begin{align}
S_{\text{SD}}=\int\md t\,\md^3y\big\{& \,i\sqrt{2}\,\tilde{\sigma}^a_{\;AB}\,\mathcal{L}_{t}A_a^{\;AB}
-H[N]  -C[N^a]  + G[t^\mu\,\sdA_\mu^{\;AB}]\,\big\} + S_{\text{SD}}^{\Delta}  \label{CanAct1}
\end{align}
with
\begin{equation} \label{SDhamiltonianConstraint}
H[N]:=\int_\Sigma\md^3y\, \frac{N}{\sqrt{q}}\,\tilde{\sigma}^a_{\;A}\,^C\,\tilde{\sigma}^b_{\;CB}\,F_{ab}^{\;AB}
\end{equation}
the Hamiltonian constraint,
\begin{equation} \label{SDvectorConstraint}
C[N^a]:=\int_\Sigma\md^3y\,i\sqrt{2}\,N^a\,\tilde{\sigma}^b_{\;AB}\,F_{ab}^{\;AB}
\end{equation}
the vector constraint, and
\begin{equation} \label{SDGaussConstraint}
G[\Lambda^{AB}]:=\int_\Sigma\md^3y \,\Lambda^{AB}\,\leftidx{^{_A}}{D}_a(i\sqrt{2}\,\tilde{\sigma}^a_{\;AB})
\end{equation}
the Gauss constraint.

In the following we will analyse in detail the canonical action (\ref{CanAct1}), and especially its boundary term $S_{\text{SD}}^{\Delta}$.


\section{Consistency conditions and differentiability}


This section has two parts. First, we state compatibility conditions on the 3+1 foliations and gauge fixing that we will impose. In the second part we explore boundary contributions to the variation of the bulk part of the canonical action. These results will be needed in the following section.

\subsection{Compatibility with 3+1 foliation and gauge fixing}

Amongst all possible foliations of the given spacetime region $M$ inducing a foliation on $\Delta$ by two-spheres, the equivalence class $[\ell]$ on $\Delta$ selects a particular subset or family of \emph{compatible foliations}. The curves of null geodesic generators may be parametrized by the time function as $x^\mu(t)$, so we require that the corresponding velocity vectors belong to the equivalence class:
\be
\ell_t^\mu:=\frac{\md x^\mu}{\md t}\in [\ell]\,.
\ee
Given any other representative $\ell^\mu\in[\ell]$, there exists then a constant $\ct$, such that 
\begin{equation} \label{evolutionField1}
\ell_t^\mu=\ct\ell^\mu\, .
\end{equation}

As discussed in  \cite{CRVwih2}, a foliation selects a particular transverse null direction $\mathbf{k}^\mu$ from the infinitely many choices in the null cone at each point $p\in\Delta$. For a given representative $\ell^\mu$, this is the unique null direction normal to $S_t=\Delta\cap\Sigma_t$, apart from $\ell^\mu$, and such that $\mathbf{k}_\mu\ell^\mu=-1$.  
Our compatibility requirement implies the 3+1 decompositions of these normals are:
\begin{equation}  \label{ellequation}
\ell^\mu=\frac{N}{\ct}\left(n^\mu+s^\mu\right)\,,
\end{equation}
\begin{equation}  \label{kequation}
\mathbf{k}^\mu=\frac{\ct}{2N}\left(n^\mu-s^\mu\right)\,.
\end{equation} 
with the lapse function $N$ assumed positive for future directed normals $\ell^\mu$, and with $s^\mu$ the `outward pointing' unit normal to $S_t$ in $\Sigma_t$. 

For non-rotating WIHs, we will choose our evolution vector field  as $t^\mu=\ell_t^\mu$.
From (\ref{ellequation}), it follows that $t^\mu = N\, \left(n^\mu+s^\mu\right)$, implying that on $\Delta$ the shift and lapse are related as $N^\mu=Ns^\mu$.

In order to have better control of the fields' variations, we will use the {\it Comoving gauge} \cite{CRVwih2}, that was also used in the initial treatments on isolated horizons \cite{ABF, ACKclassical}.
We will restrict the soldering forms and corresponding $SL(2,\mathbb{C})$ gauge transformations in a way that
$l^\mu:=i\sigma^\mu_{AA'}\om^A\bar{\om}^{A'}$   always belongs to the equivalence class $[\ell]$;  we called this the  \emph{Adapted Null gauge}.
In addition, we will require that the transverse null direction $k^\mu:=i\sigma^\mu_{AA'}\iota^A\bar{\iota}^{A'}$ coincides with $\mathbf{k}^\mu$, reducing further the gauge to $\mathbb{R}^+_\text{global}\rtimes U(1)$.  $\mathbb{R}^+_\text{global}$ refers to the global freedom of rescaling $l^\mu\to c\,l^\mu$ and $k^\mu\to c^{-1}k^\mu$ by a positive constant $c$, and $U(1)$ to rotations of $m^\mu=i\sigma^\mu_{AA'}\om^A\bar{\iota}^{A'}$ and its congugate $\bar{m}^\mu$ in the plane they span. We called this the \emph{Comoving gauge}. As usual, in the 3+1 decomposition of the self-dual action, we used the {\it time gauge} to reduce the $SL(2,\mathbb{C})$ gauge group to compact $SU(2)$. The time gauge restricts soldering forms (and gauge transformations) to satisfy $n^\mu=\sigma^\mu_{AA'}n^{AA'}$ and freezes the rescaling freedom for  $\ell$ and $\mathbf{k}$, restricting further the gauge symmetry of the WIH to $U(1)$.
We will discuss later, in the canonical framework, the (Hamiltonian) gauge symmetry of the theory on the horizon.

Solving for $n^\mu$ in (\ref{ellequation}), (\ref{kequation}) we get 
\begin{equation} \label{nequation}
n^\mu=\frac{\ct}{2N}\ell^\mu+\frac{N}{\ct}\mathbf{k}^\mu\,.
\end{equation}
and substituting the time gauge condition $n^\mu=\sigma^\mu_{AA'}\, n^{AA'}$ and  (\ref{solderingIsomorphism}) with the comoving gauge gauge conditions we get 
\begin{equation} \label{spinorNormal}
n^{AA'}=\frac{\ct}{2N}\ell^{AA'}+\frac{N}{\ct}k^{AA'}=i\left(\frac{\ct}{2N}\om^A\bar{\om}^{A'}+\frac{N}{\ct}\iota^A\bar{\iota}^{A'}\right)\,.
\end{equation}

In every point of $\Sigma$ one can identify $SU(2)$ transformations as $SL(2,\mathbb{C})$ transformations that preserve $n^{AA'}$ \cite{AshtekarLectures}. Equation (\ref{spinorNormal}) shows explicitly how the choice of compatible foliations (through $N$) determine  the $SU(2)$ subgroup of $SL(2,\mathbb{C})$.
Without loss of generality, to simplify our subsequent expressions, we will set the lapse function to a constant:
\begin{equation}
 \frac{N}{c_{(l)}}=\frac{1}{\sqrt{2}}\, .\label{N/cFixed}
\end{equation}
This gives
\be
n^{AA'}=\frac{i}{\sqrt{2}}(\om^A\bar{\om}^{A'}+ \iota^A\bar{\iota}^{A'})\, ,
\ee
leading to the standard positive definite Hermitian inner product on $\mathbb{C}^2$.

The Comoving gauge implies an expansion for the soldering form as
\begin{equation}  \label{SL2CsolderingExpansion2}
\sigma^\mu_{AA'}=-\mathbf{k}^\mu l_{AA'}- \ell^\mu k_{AA'}+\bar{m}^\mu m_{AA'}+m^\mu \bar{m}_{AA'}\,,
\end{equation}
with $m^\mu$ and $\bar{m}^\mu$ tangent to the spheres $S_t$.

Substituting (\ref{SL2CsolderingExpansion}), (\ref{spinorNormal}) and (\ref{N/cFixed}) in the defining formula (\ref{SU2solderingForm}), we obtain the gauge fixed decomposition of $\sigma^\mu_{\;AB}$ on $\Delta$
\begin{align}
\sigma^\mu_{\;AB}&=-i\sqrt{2}\,q^\mu_{\,\nu}\,\sigma^\nu_{\;AA'}\,n^{A'}_{\;\;\;B} \notag\\
&=i (\,\sqrt{2} s^\mu\,\iota_{(A}\om_{B)}-\bar{m}^\mu\,\om_A\om_B+ m^\mu\,\iota_A\iota_B\, ) .
 \label{SU2solderingExpansion}
\end{align}
The decomposition of the $SU(2)$ self-dual connection, $A_a^{AB}$, is given by the pullback of (\ref{SDAexpansion}) to $\Sigma$, with $l^\mu =\ell^\mu$ and $k^\mu =\mathbf{k}^\mu$.

\subsection{Boundary terms in the variations of the constraints}

In the next section we will construct  well defined Hamiltonians and the corresponding HVF $X_H$, that also describe consistent dynamics on $\Delta$. The first step in the  construction of the WIH components of $X_H$ is the analysis of boundary terms that appear in the variations of the constraints. We need the expressions from the previous subsection in order to calculate these variations.

The boundary contribution from variation of the Hamiltonian constraint (\ref{SDhamiltonianConstraint}) is
\begin{equation} \label{SDhamiltonianVar}
\delta H[N]|_{S_\Delta}=\oint_{S_\Delta} \md S_a\,\frac{2N}{\sqrt{q}}\,\tilde{\sigma}^{[a|}_{\;A}\,^C\,\tilde{\sigma}^{|b]}_{\;CB}\,\delta A_{b}^{\;AB}\, ,
\end{equation}
where $S_\Delta = \Sigma\cap\Delta$.

The boundary contribution from variation of the vector constraint (\ref{SDvectorConstraint}) is
\begin{equation} \label{SDvectorVar}
\delta C[N^a]|_{S_\Delta}=i\sqrt{2}\oint_{S_\Delta} \md S_a\,2Ns^{[a}\,\tilde{\sigma}^{b]}_{\;AB}\,\delta A_{b}^{\;AB} \, .
\end{equation}
where we have taken into account that on $\Delta$, $N^a=Ns^a$.

Expressions  (\ref{SDhamiltonianVar}) and (\ref{SDvectorVar})  are $SU(2)$-gauge invariant. We can therefore  substitute expansion (\ref{SDAexpansion})  for the connection  in the comoving gauge, and the gauge fixed expression (\ref{SU2solderingExpansion}) for the soldering form, to find reduced expressions for these variations.
We also need the expression for
\be
\md S_a=\frac{1}{2}\epsilon_{abc}\md x^b\wedge\md x^c=\frac{s_a}{\sqrt{q}}\,\sqrt{h}\,\md^2\Theta=\frac{s_a}{\sqrt{q}}\,\leftidx{^2}{\epsilon} \, .
\ee
Then, the boundary term   (\ref{SDhamiltonianVar}) takes the form
\be
\delta H[N]|_{S_\Delta} =
 \sqrt{2}\oint_{S_\Delta} N\, \bigl(\bar{m}^b \om_{(A}\om_{B)} + m^b \iota_A\iota_B\bigr) \,\delta A_{b}^{\;AB}\,\leftidx{^2}{\epsilon}= -\sqrt{2}\oint_{S_\Delta}N\, m^b\,\delta\U_b\,\leftidx{^2}{\epsilon}\, ,
\ee
where we have used that for a NEH, in Adapted Null Gauge, we have
$\Y_\pbi{\mu}=0$ \cite{CRVwih2},  
so that $\bar{m}^b\,\delta\Y_b=0$.

Similarly,  boundary term  (\ref{SDvectorVar}) is of the form
\be
\delta C[N^a]|_{S_\Delta}
= \sqrt{2}\oint_{S_\Delta} N\, \bigl(\bar{m}^b \om_{(A}\om_{B)} - m^b \iota_A\iota_B\bigr) \,\delta A_{b}^{\;AB}\,\leftidx{^2}{\epsilon} 
=\sqrt{2}\oint_{S_\Delta}N\, m^b\,\delta\U_b\,\leftidx{^2}{\epsilon}\, .
\ee
As a result, we obtain
\be\label{Var0}
\delta H[N]|_{S_\Delta} + \delta C[N^a]|_{S_\Delta} = 0 \, .
\ee
Note that this is a remarkable result (and true for general lapse $N$). While in the bulk, the lapse $N$ and shift vector $N^a$ are independent, generating ``bubble time evolution", requiring that time evolution is compatible with all the structures at the horizon forces one to consider lapse and shifts that work in `tandem'. Thus, the combined contribution to the variation of these two terms in the canonical Hamiltonian vanishes. This also means that if there is any boundary term one has to add to the canonical Hamiltonian to make it differentiable ({\it a la} Regge-Teitelboim), it would be to Gauss' law, in stark contrast to the case in spatial infinity with asymptotically flat boundary conditions \cite{CorichiReyes}.
Finally, the variation of the Gauss constraint has indeed a boundary term
\begin{equation}\label{VarGauss}
\delta G[t^\mu\,\sdA_\mu^{\;AB}]|_{S_\Delta} = i\sqrt{2}\oint_{S_\Delta} \md S_a\, (t^\mu\,\sdA_\mu^{\;AB})\,
\delta\tilde{\sigma}^a_{\;AB} = \oint_{S_\Delta} t\cdot (\omega +V)\,\delta \,\leftidx{^2}{\epsilon}\, ,
\end{equation}
where we have also used that in the Adapted Null gauge 
$\W_\pbi{\mu}=\omega_\pbi{\mu}$ and 
$\V_\pbi{\mu}=V_\pbi{\mu}$. 

In the following section we will show in detail  that, due to (\ref{Var0}), the boundary equations of motion are obtained only from the Gauss constraint, in contrast to the bulk equations of motion. 
\section{Hamiltonians and boundary phase spaces}

This is the main section of this manuscript. The objective is {to} explore the boundary term in the canonical action, when considering the WIH boundary conditions. We shall see that the remaining terms, can be interpreted as either a contribution to the ``kinetic'' term, or as a boundary contribution to the Hamiltonian. Depending on the choice, we shall have different canonical descriptions of the dynamics of the horizon. This section has four parts, in each of which we consider each of the four possible choices for the boundary terms.

Let us now turn our attention to 
the boundary term $S_{\text{SD}}^{\Delta}$ in the canonical  action. Using the expansions (\ref{SDAexpansion}) and (\ref{SU2solderingExpansion}), it can be written as
\be
S_{\text{SD}}^{\Delta} =\int_\Delta\,\frac{1}{N}\, (t^\mu\,\sdA_\mu^{\;AB})\,(i\sqrt{2}\,{\sigma}^a_{\;AB}\,\tilde{e}^\nu_a) \ell_\nu\leftidx{^2}{\epsilon}\,\md\lambda  
=-\int \md t \oint_{S_\Delta}t\cdot (\W + \V)\,\leftidx{^2}{\epsilon} 
\label{GaugeFixedSDboundary}
\ee
where we have also used  $\sqrt{2}\, \tilde{e}^\nu_a \ell_\nu  = s_a$ and $\lambda = c_{(l)}t$.  In the Adapted Null Gauge, we can finally obtain
\be
S_{\text{SD}}^{\Delta}=-\int \md t \oint_{S_\Delta} t\cdot (\omega +  V )\,\leftidx{^2}{\epsilon} \, .\label{GaugeFixedSDboundary1}
\ee
Note that this boundary term is not invariant under the residual $U(1)$ gauge transformations on $\Delta$, namely rotation of $m$ and $\bar{m}$ preserving  $\ell$ and $\mathbf{k}$: 
$\ell\to \ell$, $\mathbf{k}\to \mathbf{k}$,  $m\to e^{i\theta}m$,   $\bar{m}\to e^{-i\theta}\bar{m}$, for 
$\theta\in\mathbb{R}$. Under this transformation $\omega$ is invariant, but $V$ transforms as a $U(1)$ connection: $V\to V+i\md\theta$.

We shall now show that $S_{\text{SD}}^{\Delta}$ can be seen either as a kinetic term in the canonical action or as a boundary term in the corresponding Hamiltonian, or a mixture of both. This statement might sound strange since the boundary term (\ref{GaugeFixedSDboundary1}) contains two contributions, both of which contain some part of the self-dual connection. Since the action principle was written under the assumption that the connection shall play the role of configuration variable (as is the case in gauge theories without boundaries), one might argue that neither term contains a time derivative of any of the canonical variables, and thus, it cannot be a kinetic term. How can one view either (or both) terms as kinetic? It is not difficult to imagine that one can complete that feat by introducing {\it new} degrees of freedom at the boundary. This is precisely what is done in electrodynamics when one introduces the magnetic potential as a ``new" degree of freedom, and the magnetic field becomes a derived object. These degrees of freedom are not entirely new, since they are potentials for the already existing ones. 
Here, the strategy is to view $\omega$ and/or $V$ as arising from a potential. Thus, we have four possibilities. The first one is to consider both terms as arising from a potential. Then, we shall have two cases in which only one of the terms is reinterpreted. An finally, the fourth case is the ``original" one in which one does {\it not} introduce a potential and simply interprets (\ref{GaugeFixedSDboundary1}) as a contribution to the Hamiltonian.

In what follows, we will explore the consistency of the Hamiltonian formulation of the theory based on all these possibilities.

\subsection{Case I}

As we discussed before, in this part we shall explore the possibility that both $\omega$ and $V$ arise from some potential.
With this idea in mind, let us introduce $\psi_R$ and $\psi_I$, that we should call the real and imaginary potentials, respectively, such that
\begin{eqnarray}
\mathcal{L}_t{\psi_R} &=& t\cdot\md\psi_R\WIHeq t\cdot\omega\, ,\label{real_psi}\\
\mathcal{L}_t{\psi_I} &=& t\cdot\md\psi_L\WIHeq t\cdot V\, .\label{imaginary_psi}
\end{eqnarray}
In this first case we will consider a complex potential of the form $\psi := \psi_R + \psi_I$ (recall that $V$ is already imaginary), and its corresponding canonical momentum $\tilde{\pi}$ as our boundary degrees of freedom. As we shall see in detail, the formalism will tell us what $\tilde{\pi}$ corresponds to.

\subsubsection{Symplectic structure and the Hamiltonian}
The idea here is to rewrite the boundary term as
\begin{equation}
 S_{\text{SD,1}}^{\Delta}=-\int \md t \oint_{S_\Delta} (\mathcal{L}_{t}\, \psi )\,\leftidx{^2}{\epsilon} \, .
\end{equation}
In this case, it contributes completely to the kinetic term in the canonical action, that indicates the existence of boundary degrees of freedom on the horizon.
As a result, the canonical action takes the following form
\begin{equation}
S_{\text{SD,1}}=\int\md t\,\big\{ \int_{\Sigma}\md^3y \,i\sqrt{2}\,\tilde{\sigma}^a_{\;AB}\,\mathcal{L}_{t}A_a^{\;AB} - \oint_{S_{\Delta}} \md^2\Theta \sqrt{h} \,\mathcal{L}_{t} \psi -  H_{\text{SD,1}}\,\big\}
\end{equation}
where 
the candidate for the canonical Hamiltonian is given by
\begin{equation}
  H_{\text{SD,1}}= H[N]  +C[N^a]  - G[t^\mu\,\sdA_\mu^{\;AB}]\, .
\end{equation}
Since $S_{\text{SD, 1}}$ contains time derivative of $\psi$ on $\Delta$, the 
kinetic term  has a contribution from the boundary. In this case we have that the $\Gamma_{\rm Bulk}$ part of the phase space is parametrized by the coordinates $(A_a^{\;AB},\tilde{P}^a_{\;AB})$   and $\Gamma_{\rm Bound}$ is parametrized by $(\psi , \tilde{\pi})$, where
\begin{eqnarray}
 \tilde{P}^a_{\;AB} &=& i\sqrt{2}\,\tilde{\sigma}^a_{\;AB}\, .\\
 \tilde{\pi} &=& -\sqrt{h}\, ,
\end{eqnarray}
Note that $\tilde{\pi}_I = \tilde{\pi}_R = \tilde{\pi}$, since only the combination $\psi =\psi_R + \psi_I$ appears in the kinetic term, while $\psi^*=\psi_R - \psi_I$ is not dynamical, leaving us with two canonical degrees of freedom on the boundary $\Delta$.
The corresponding symplectic structure has a boundary term and is of the form
\be\label{SymplStructSD1}
\Omega_1 =\int_\Sigma\md^3x\,\,\ed\tilde{P}^a_{\;AB}\dwedge \ed A_a^{\;AB} +  \oint_{S_\Delta}\md^2\Theta\, 
\,\ed\tilde{\pi}\dwedge\ed\psi\, .
\ee
In the canonical framework the equations of motions are given by 
\begin{equation}
\ed H_{\text{SD,1}} (Y) =\Omega_1 (Y ,X_H)\, , \label{HamEqs}
\end{equation}
where $X_H$ is the corresponding Hamiltonian vector field,
whose components in the bulk are $(X_{H,A}, X_{H,\tilde{P}})$ and on the boundary $(X_{H,\psi}, X_{H,\tilde{\pi}})$. 
The bulk part of (\ref{HamEqs}) gives the equations of motion in the bulk, while
the boundary term  defines the boundary components of the Hamiltonian vector field.

Here, we are especially interested in the boundary contributions in (\ref{HamEqs}). Using the previously obtained expressions (\ref{Var0}) and (\ref{VarGauss}), it follows that
\be\label{VarH1}
\delta H_{\text{SD,1}}|_{S_\Delta} = -\oint_{S_\Delta} t\cdot (\omega +V)\delta \,\leftidx{^2}{\epsilon}\, .
\ee
We can also rewrite it in terms of en exterior derivative on the phase space,
\be\label{dH1}
\ed H_{\text{SD,1}}(Y)|_{S_\Delta} = \oint_{S_\Delta}\md^2\Theta\  [t\cdot (\omega +V)]\, \ed\tilde{\pi}(Y)=\oint_{S_\Delta}\md^2\Theta\  [t\cdot (\omega +V)]\, Y_{\tilde{\pi}}\, ,
\ee
where we have used the property $\ed\tilde{\pi} (y)\left( \frac{\delta}{\delta \tilde{\pi}(y')} \right) =\delta^2(y,y')$. 
On the other hand, from (\ref{SymplStructSD1}) it follows that the boundary contribution to the symplectic structure is
\begin{equation}
 \Omega_1 (Y,X)|_{S_\Delta} = \oint_{S_\Delta}\md^2\Theta\, (X_{\psi}Y_{\tilde{\pi}}-Y_{\psi}X_{\tilde{\pi}})\, .\label{BoundarySS1}
\end{equation}
From (\ref{dH1}) and (\ref{BoundarySS1}) we can read off the
the boundary components of the corresponding HVF, $X_{\text{H,1}}$, 
\begin{eqnarray}
 X_{H1, \tilde{\pi}} = 0 \ \ \ &\Rightarrow & \ \ \ \mathcal{L}_t\leftidx{^2}{\epsilon}=0\, ,\\
 X_{H1,\psi} = t\cdot\omega + t\cdot V\ \ \ &\Rightarrow & \ \ \ \mathcal{L}_t\psi = t\cdot\omega + t\cdot V\, .
\end{eqnarray}
The first equation is a consequence of the WIH boundary conditions and the second one is the definition of the potential $\psi$. In that sense,  we have obtained a consistent evolution of the boundary degrees of freedom.

Nevertheless, $H_{\text{SD,1}}$ vanishes on the constraint surface, and thus it does not contain the information about the mass of the WIH.  In the following we shall see that, given the freedom in the election of the symplectic structure on the boundary, we can find a Hamiltonian with a non-vanishing boundary term. Before that, let us analyze the generators of gauge and diffeomorphism transformations in the current setup.


\subsubsection{Gauss constraint}

Here, we want to analyse the residual gauge symmetry on a WIH. For that we need to find out the form of the boundary components of the corresponding HVF. Previously, we have restricted the residual internal gauge symmetry on a WIH to $U(1)$. Now, we want to  further explore this gauge reduction from the canonical framework. We start with the Gauss constraint that is given by 
\begin{equation} \label{SDGaussConstraint1}
G[\Lambda^{AB}]:=\int_\Sigma\md^3y \,\Lambda^{AB}\,\leftidx{^{_A}}{D}_a\tilde{P}^a_{\;AB}
=-\int_\Sigma\md^3y \, (\leftidx{^{_A}}{D}_a\Lambda^{AB})\,\tilde{P}^a_{\;AB}
+\oint_{S_\Delta}\, \md S_a\, \Lambda^{AB}\,\tilde{P}^a_{\;AB}\, .
\end{equation}
The general smearing function $\Lambda^{AB}$ can be expanded as
\be
\Lambda^{AB}=i\Lambda\, \iota^{(A}\om^{B)}+\Omega\,\om^{A} \om^{B} + \Theta\, \iota^A \iota^B\, ,
\ee
and it has to be traceless ${\Lambda^A}_A=0$ and anti-hermitian $\Lambda^{AB} =-\bar{\Lambda}^{BA}$. These conditions imply that  
\be
\Lambda^{AB}=i\Lambda\, \iota^{(A}\om^{B)}+\Omega\,\om^{A} \om^{B} - \bar{\Omega}\, \iota^A \iota^B\, ,\label{G2}
\ee
with $\Lambda\in\mathbb{R}$  and $\Omega\in\mathbb{C}$.

The boundary term in (\ref{SDGaussConstraint1}) can be rewritten using the expansion of $\tilde{\sigma}^a_{\;AB}$ in the Comoving Gauge   (\ref{SU2solderingExpansion}) as
\begin{equation}\label{GCBT}
\oint_{S_\Delta}\, \md S_a\, \Lambda^{AB}\,\tilde{P}^a_{\;AB} = -2 \oint_{S_\Delta}\,
\Lambda^{AB}\, \iota_{(A}\om_{B)}\, \leftidx{^2}{\epsilon}
=\oint_{S_\Delta}\, i\Lambda\,  \leftidx{^2}{\epsilon}\, .
\end{equation}

The  vector field $X_G$, that defines the infinitesimal gauge transformations, is given by
\begin{equation}
\ed G (Y) = \Omega_1 (Y, X_G)\, ,\label{GCVFEq}
\end{equation}
where $Y$ is an arbitrary vector in $T\Gamma$. 
Again, we will only consider the WIH contribution in (\ref{GCVFEq}). Taking into account (\ref{SDGaussConstraint1}) and (\ref{GCBT}) we obtain the form of the boundary gauge transformation
\begin{equation}
 \delta_g\, \tilde{\pi} := X_{G,\tilde{\pi}} =  0
\end{equation}
\begin{equation}
 \delta_g\psi := X_{G,\psi} = -i \Lambda\, .
\end{equation}
Since, $\omega =\md\psi_R + \tilde{\omega}$ and $V =\md\psi_I + \tilde{V}$  (the explicit form of $\tilde{\omega}$ and $\tilde{V}$ is given in the Appendix A), the Gauss constraint generates on a WIH gauge transformations: $\psi_R\to\psi_R$ and $\psi_I\to\psi_I -i\Lambda$, that corresponds to $U(1)$ transformations: $\omega\to\omega$ and
 $V\to V-i\,\md\Lambda$. Note that these transformations preserve the area of $S_\Delta$.

The $X_G$, defined by the  previous expressions, is  a degenerate direction of the symplectic structure, since the pullback of the symplectic structure to the submanifold defined by the Gauss constraint, $\Omega_g$, is degenerate. Effectively,  
\begin{equation}
\Omega_g (Y,X_G) = 0\ \ \ \Leftrightarrow\ \ \ \ed G[\Lambda^{AB}](Y)\approx 0\, .
\end{equation}
The R.H.S. is always true for any vector $Y$ tangent to the Gauss constraint's surface. 
It should also be true, for any vector tangent to diffeomorfism and Hamiltonian constraints. For that to happen all the constraints should be first class.


\subsubsection{Spatial diffeomorphism constraint}

Let us see what are the residual diffeomorphism symmetries on a WIH, in this case.
It is well known that the generator of spatial diffeomorphisms on $\Sigma$ is a linear combinations of the vector (\ref{SDvectorConstraint}) and the Gauss (\ref{SDGaussConstraint}) constraints \cite{ThiemannBook}, and is of the form
\be
D[N^a]=\int_\Sigma\md^3y\,N^a\,(\,\tilde{P}^b_{\;AB}\,F_{ab}^{\;AB}-
A_a^{\;AB}\,\leftidx{^{_A}}{D}_b\tilde{P}^b_{\;AB}\,)\, .
\ee
To analyse tangential diffeomorphisms on  WIH we need to calculate $\delta D[N^a]\vert_{S_\Delta}$, and the  boundary components of the corresponding HVF in the case when the shift is tangential to $S_\Delta$, 
\be
N^a = \mu\, m^a + \bar{\mu}\, \bar{m}^a\, ,
\ee
where $\mu$ is an arbitrary functions on $S_\Delta$. 

The corresponding HVF $X_D$ is determined from
\be
\ed D (Y)=\Omega (Y, X_D)\, .
\ee
Boundary components of $X_D$ determine residual diffeomorphisms on WIH. 
The boundary term in $\delta D[N^a]$ is of the form
\be
\delta D[N^a]\vert_{S_\Delta} 
=
-\oint_{S_\Delta}\md^2\Theta\, \delta \bigl[ N^a (\omega_a + V_a) \, \tilde{\pi}\bigr]:=
\delta \bar{D}[N^a]\vert_{S_\Delta}%
+\delta \hat{D}[N^a]\vert_{S_\Delta}\, ,\label{deltaD}
\ee
where, using the decompositions (\ref{Decomp1})-(\ref{Decomp2}), we have introduced
\begin{eqnarray}
\delta \bar{D}[N^a]\vert_{S_\Delta} &:=&-\oint_{S_\Delta}  \md^2\Theta\,  N^b\, \delta\bigl[(\nabla_b\psi ) \, \tilde{\pi}\bigr]\, ,\label{D1}\\
\delta \hat{D}[N^a]\vert_{S_\Delta} &:=-&\oint_{S_\Delta} \md^2\Theta\,
N^b\,\delta [(\tilde\omega_b +\tilde{V}_b)\, \tilde{\pi}]\, .\label{D2}
\end{eqnarray}
As we have defined in  (\ref{gradconfrontera}) the boundary contribution in the variation of any allowed functional should only depend on boundary degrees of freedom, in this case it should  be of the form
\be
\delta F\vert_{S_\Delta} =\oint_{S_\Delta}\md^2\Theta\, \left(\, \frac{\delta F}{\delta\psi} \delta \psi + \frac{\delta F}{\delta\tilde{\pi}} \delta \tilde{\pi}\, \right) \, .\label{VarF}
\ee
We see that the $\delta \bar{D} [N^a]\vert_{S_\Delta}$ is of the form given in (\ref{VarF}), but $\delta \hat{D} [N^a]\vert_{S_\Delta}$ is not and it does not vanish either. 
As a consequence $D [N^a]$ is an allowed funcional
only in the case when $N^a\WIHeq 0$. Then $\delta D [N^a]\vert_{S_\Delta}=0$, as a result, the boundary components of $X_D$ vanish, that means that
the tangential diffeomorphisms are not gauge transformations on the boundary.  In the Appendix B we will show that in the special case, when we restrict to spherically symmetric configurations and the phase space with fixed area of $S_\Delta$, as in \cite{ACKclassical}, 
$D[N^a]$ generates
tangential diffeomorphisms on $S_\Delta$.

To end this part, let us note that we can add a boundary term to $D[N^a]$ as
\be\label{Dimproved}
\tilde{D}[N^a]=D[N^a]+\oint_{S_\Delta}\md^2\Theta\,  \bigl[ N^a (\tilde{\omega}_a + \tilde{V}_a) \, \tilde{\pi}\bigr]\, .
\ee
such that $\delta D[N^a]\vert_{S_\Delta}= \delta \bar{D}[N^a]\vert_{S_\Delta}$. 
The boundary component of HVF corresponding to $\tilde{D}$ is obtained from
\begin{equation}
\ed \tilde{D} (Y)\vert_{S_\Delta} = \Omega (Y, X_{\tilde{D}})\vert_{S_\Delta}\, ,\label{HVFDiffTang1}
\end{equation}
leading to
\begin{equation}
\oint_{S_\Delta}\md^2\Theta\, \bigl[-\nabla_a(N^a\tilde{\pi})\, Y_{\psi} + N^a(\nabla_a\psi )\, Y_{\tilde{\pi}}\bigr]  = \oint_{S_\Delta}\md^2\Theta\, (X_{\tilde{D},\psi}Y_{\tilde{\pi}}-Y_{\psi}X_{\tilde{D},\tilde{\pi}})\, .
\end{equation}
Then, it follows that
\begin{eqnarray}
X_{\tilde{D},\tilde{\pi}} =  \nabla_a (N^a\tilde{\pi})\ \ \ &\Rightarrow& \ \ \
\tilde{\delta}\sqrt{h}=\nabla_a(N^a\sqrt{h})=\mathcal{L}_{\vec{N}}\sqrt{h}\, ,\notag\\
X_{\tilde{D},\psi} = N^a\nabla_a\psi \ \ \ &\Rightarrow& \ \ \ \tilde{\delta}\psi = N^a\nabla_a\psi = \mathcal{L}_{\vec{N}}\psi\, ,
\end{eqnarray}
and the 'improved' generator indeed generate tangential diffeomorphisms on $S_\Delta$. 
This corresponds precisely to the Regge-Teitelboim strategy of constructing Hamiltonians to generate specific symplectomorphism on phase space. In this case, the generator of horizon diffeomorphisms, that is a true (non-vanishing on shell) Hamiltonian function, and {\it not} a constraint. Note that this is very similar to the result obtained in the covariant formalism in \cite{ABL}. 
To summarize this choice. When we take both terms as the boundary term of the canonical action to be dynamical, namely to contribute to the symplectic structure, we do get a consistent description but we do not recover the (expected) mass of the black hole as a boundary term. Let us now consider cases in which only one of the terms are to be regarded as dynamic.


\subsection{Case II}

In this case we will consider $(\psi_I ,\tilde{\pi})$ as  boundary degrees of freedom and leave the first term in (\ref{GaugeFixedSDboundary1}) in the form of a boundary term in the Hamiltonian. This approach is an extension of the results of \cite{ABF} to WIHs. We will show that in this case, the resulting  boundary term in the Hamiltonian shall correspond to the mass of a WIH. 

Since  $\kappa_{(t)}:=t\cdot \omega$ , we can rewrite the boundary term in the canonical action (\ref{GaugeFixedSDboundary1}) as
\begin{equation}\label{SboundarySP}
 S_{\text{SD,2}}^{\Delta} =-\int \md t\oint_{S_\Delta} \, (\kappa_{(t)}\,  +  \mathcal{L}_t\psi_I ) \,\leftidx{^2}{\epsilon} \, .
\end{equation}
The last integral over $S_\Delta$ in the previous expression is the symplectic potential on the horizon, that gives rise to the symplectic structure on the horizon. 

The canonical action is now of the form
\begin{equation}
S_{\text{SD,2}}=\int\md t\,\big\{ \int_{\Sigma}\md^3y \,i\sqrt{2}\,\tilde{\sigma}^a_{\;AB}\,\mathcal{L}_{\bar{t}}A_a^{\;AB} - \oint_{S_{\Delta}} \md^2\Theta \sqrt{h} \,\mathcal{L}_{\bar{t}} \psi_I -  H_{\text{SD,2}}\,\big\}
\end{equation}
where the candidate for the canonical Hamiltonian is given by
\begin{equation}
  H_{\text{SD,2}}=H_{\text{SD,1}}+\oint_{S_\Delta} \, \kappa_{(t)}\, \leftidx{^2}{\epsilon} \, ,
\end{equation}
and the symplectic structure in this case is of the form
\be\label{SymplStructSD2}
\Omega_2 =\int_\Sigma\md^3x\,\,\ed\tilde{P}^a_{\;AB}\dwedge \ed A_a^{\;AB} +  \oint_{S_\Delta}\md^2\Theta\, 
\,\ed\tilde{\pi}\dwedge\ed\psi_I\, .
\ee
Now, only $\psi_I$ and $\tilde{\pi}$ contribute to $\Omega_2$, and are to be regarded as the boundary degrees of freedom.

In order to  calculate the boundary contribution to (\ref{HamEqs}), let us first analyze the variation of the Hamitonian, that is given by
\begin{equation}
\delta H_{\text{SD,2}}\vert_{S_\Delta} 
 =\oint_{S_\Delta}\md^2\Theta\, \big[\, -\delta\kappa_{(t)}\,\tilde{\pi}  +  (t\cdot V)\, \delta\, \tilde{\pi}\, \bigr]
 \, .\label{VarHamBound}
\end{equation}
Comparing it to the variation of an allowed function, we can see that the first term in (\ref{VarHamBound}) should vanish or $\delta\kappa_{(t)}$ should be a linear combination of  $\delta\psi$ and $\delta\tilde{\pi}$.

The corresponding HVF, $X_{\text{H2}}$ is obtained from
\begin{equation}
 \ed H_{\text{SD,2}}(Y) =\Omega_2 (Y, X_{\text{H2}})\, , \label{dHYII}
\end{equation}
Then, from (\ref{SymplStructSD2}) and (\ref{VarHamBound}), it follows that the boundary contribution to the previous expression is given by
\be
\oint_{S_\Delta}\md^2\Theta\, \big[\, -\tilde{\pi}\, Y_{\kappa_{(t)}} +  (t\cdot V)\, Y_{\tilde{\pi}}\, \bigr]
=\oint_{S_\Delta}\md^2\Theta\, (X_{H,\psi_I}Y_{\tilde{\pi}}-X_{H,\tilde{\pi}}Y_{\psi_I})\, ,
\ee
where we have introduced the notation $Y_{\kappa_{(t)}}:=\delta\kappa_{(t)}$. 
In order to have the consistent description of the evolution on the boundary we should have $Y_{\kappa_{(t)}} = A  Y_{\psi_I}+ B Y_{\tilde{\pi}}$, with $A$ and $B$ functions on $S_{\Delta}$. Then, it follows that $X_{H,\psi_I}=B\sqrt{h}+t\cdot V$ and $X_{H,\tilde{\pi}}=-A$, leading to
$\mathcal{L}_l\psi_I = B\sqrt{h}+t\cdot V$ and $\mathcal{L}_l\sqrt{h}=A$. We see that only when $A=B=0$, we can obtain the correct boundary condition  $\mathcal{L}_l\sqrt{h}\WIHeq 0$, and also $\mathcal{L}_l\psi_I = t\cdot V$. 

So, we are led to the conclusion that we have to eliminate the first term 
in (\ref{VarHamBound}). That can be done by adding a counterterm to the Hamiltonian, $H_{\text{CT}}$, such that
\be\label{VarHT}
\delta (H_{\text{SD,2}}+H_{\text{CT,2}})\vert_{S_\Delta}=\oint_{S_\Delta}\,\md^2\Theta\,  (t\cdot V)\, \delta\, \tilde{\pi}\, ,
\ee
that implies that
\be\label{VarCT}
\delta H_{\text{CT,2}}= \oint_{S_\Delta}\, \md^2\Theta\, \delta\kappa_{(t)}\, \tilde{\pi}\, .
\ee
In \cite{CRVwih2} and \cite{GJreview} it has been shown that $(h_{ab}, \Omega_a, \Xi_{ab}, \kappa_{(t)})$ are the free independent parameters on a given sphere $S_\Delta$, completely determining the WIH geometry. 
The \emph{H\'a\'{\j}i\v{c}ek 1-form} \cite{Hajicek} $\Omega_\mu:=h^\nu_{\;\mu}\omega_\nu$ is the projection  of the rotation 1-form to the sphere and which also coincides with its pullback $\omega_a$ on $T_p\mathcal{S}_{\Delta}$: $\omega_a=\Omega_a$. The \emph{transversal deformation rate}
$\Xi_{\mu\nu}:=h^\sigma_{\;\;\mu}h^\rho_{\;\;\nu}\nabla_\sigma k_\rho$ is the analog for $k^\mu$ of the second fundamental form (\ref{ThetaDef}).
$\Omega_a$ and $\Xi_{ab}=\nabla_ak_b$ are completely unrestricted and $\Omega_\mu$ and $\Xi_{\mu\nu}$ may be determined from them  on all of $\Delta$ using certain constraint equations for $\mathcal{L}_\ell\,\Omega_\mu$ and $\mathcal{L}_\ell\,\Xi_{\mu\nu}$ whose exact form is given in \cite{GJreview, ABLgeometry}.

It follows then that a general counterterm must have the form
\begin{equation}
H_{\text{CT}}=\oint_{S_\Delta} F(h_{ab}, \Omega_a, \Xi_{ab}, \kappa_{(t)})\,\leftidx{^2}{\epsilon}\, ,
\end{equation}
and its variation is
\begin{equation}\label{GenVarCT}
\delta H_{\text{CT}}=\oint_{S_\Delta}\left\{\left(\vd{F}{h_{ab}}\delta h_{ab}+\vd{F}{\Omega_a}\delta \Omega_a+\vd{F}{\Xi_{ab}}\delta\Xi_{ab}+\vd{F}{\kappa_{(t)}}\delta\kappa_{(t)}\right)\,\leftidx{^2}{\epsilon} 
+F\,\delta\,\leftidx{^2}{\epsilon}\right\}\, .
\end{equation}
Assuming this variation cancels the unwanted boundary term in (\ref{VarHamBound}), it
implies that variations $(\delta h_{ab}, \delta \Omega_a, \delta\Xi_{ab}, \delta \kappa_{(t)})$ are not fully independent. Namely, a priori, $(h_{ab}, \Omega_a, \Xi_{ab}, \kappa_{})$ are independent, so are their variations. We can consider directions on phase space such that $\delta h_{ab}=0$ (and hence $\delta\leftidx{^2}{\epsilon}=0$) and $\delta\kappa_{(t)}=0$.
This implies
\begin{equation}
\Par{F}{\Omega_a}=\Par{F}{\Xi_{ab}}=0\, .
\end{equation}

Now, from (\ref{VarCT}) and (\ref{GenVarCT}), it follows
\be\label{VarCT1}
 \delta H_{\text{CT}}=\oint_{S_\Delta}\left\{\left(\vd{F}{h_{ab}}\delta h_{ab}+\vd{F}{\kappa_{(t)}}\delta\kappa_{(t)}\right)\,\leftidx{^2}{\epsilon} 
+F\,\delta\,\leftidx{^2}{\epsilon}\right\}= -\oint_{S_\Delta}\, \delta\kappa_{(t)}\,\leftidx{^2}{\epsilon}\, .
\ee
Then, taking variations in directions where $\delta\kappa_{(t)}=0$, gives the condition
\be
\oint_{S_\Delta}\left\{\left(\vd{F}{h_{ab}}\delta h_{ab}\right)\,\leftidx{^2}{\epsilon} 
+F\,\delta\,\leftidx{^2}{\epsilon}\right\}=0\, ,
\ee
that implies $F$ should only depend on the metric $h_{ab}$ through the area, $F(a_\Delta,\kappa_{(t)})$. Since $a_\Delta$ and $\kappa_{(t)}$ are constant on $S_{\Delta}$, from (\ref{VarCT1}), we obtain, for the general variations, that
\be
\left(\Par{F}{\kappa_{(t)}}+1\right)a_\Delta\delta\kappa_{(t)}+\left(a_\Delta\Par{F}{a_\Delta}+F\right)\delta a_\Delta =0\, .
\ee
This in turn implies that $\kappa_{(t)}$ must be a function of the area  $a_\Delta$. (Indeed, if $\delta\kappa_{(t)}$ and $\delta a_\Delta$ were independent, we would need $\Par{F}{\kappa_{(t)}}+1=0$, which implies $F(a_\Delta,\kappa_{(t)})=-\kappa_{(t)}+C(a_\Delta)$. But then $a_\Delta\Par{F}{a_\Delta}+F=a_\Delta\Par{C}{a_\Delta}+C-\kappa_{(t)}$ cannot vanish in general since $C$ is independent of $\kappa_{(t)}$.)

So in the canonical formulation, we recover the result of the covariant Hamiltonian analysis of \cite{AFK}, that the surface gravity must only be a function of $a_\Delta$, 
\begin{equation}
\kappa_{(t)}=\kappa_{(t)}(a_\Delta)\, .
\end{equation}
As a consequence, the counterterm is also a function of area only:
 \be\label{area}
H_{\text{CT}}=\oint_{S_\Delta} F(a_\Delta)\,\leftidx{^2}{\epsilon}= F(a_\Delta)\,a_\Delta\, .
\ee
and therefore, from (\ref{VarCT1}) and (\ref{area}), we obtain
\be
 \delta H_{\text{CT}}=(\frac{\md {F}}{\md a_\Delta}a_\Delta + F)\,\delta a_\Delta= - \frac{\md\kappa_{(t)}}{\md a_\Delta}\, a_\Delta\, \delta a_\Delta\, ,
\ee
that leads to  a differential equation for $F(a_\Delta)$:
\begin{equation} \label{DiffeqCounter}
a_\Delta\,\frac{\md F}{\md a_\Delta}
+F=-a_\Delta\frac{\md\kappa_{(t)}}{\md a_\Delta}\, .
\end{equation}
It is standard to fix
\begin{equation} \label{kappaStatic}
\kappa_{(t)}=\frac{1}{2R_\Delta}=\sqrt{\pi}\,a_\Delta^{-1/2}\, ,
\end{equation}
where $a_\Delta=4\pi R_\Delta^2$. In this way on a static solution $t^\mu$ coincides with the Killing vector field on $\Delta$.
This choice implies the choice of $t^\mu$, so that we can define the horizon mass as $M_\Delta = E^t_\Delta$, for any point of the 
phase space, not only for static spacetimes \cite{CRVwih2}. 

The general solution to (\ref{DiffeqCounter}) is $F=\kappa_{(t)}+ C a_\Delta^{-1}$, where $C$ is an arbitrary constant, so that
\be\label{CT2}
H_{\text{CT,2}}=\int_{S_\Delta}\, \kappa_{(t)} \,\leftidx{^2}{\epsilon} + C \, .
\ee
Without the loss of generality, we can  fix $C=0$, and obtain  
a well defined Hamiltonian $H_{2}=H_{\text{SD,2}}+H_{\text{CT}}$ as
 \begin{equation}\label{TotalHamCT}
 H_{2}=\int_\Sigma\md^3y\, \big\{ 
H[N] + C[N^a]  - G[t^\mu\,\sdA_\mu^{\;AB}]\,\big\} + 2\oint_{S_\Delta} \, \kappa_{(t)}\,\leftidx{^2}{\epsilon} \, .
\end{equation}
The boundary term corresponds to the Smarr expression for the mass of a non-rotating black hole \cite{Smarr}. 


Let us analyse now the canonical description of the evolution generated by $H_2$ in the phase space.
The equations of motions are given by 
\begin{equation}
\ed H_{2} (Y) =\Omega (Y ,X_{H2})\, , \label{HamEqs1}
\end{equation}
where $X_{H2}$ is the corresponding Hamiltonian vector field,
whose components in the bulk are $(X_{H2,A}, X_{H2,\tilde{P}})$ and on the boundary $(X_{H2,\psi}, X_{H2,\tilde{\pi}})$. 
The bulk part of (\ref{HamEqs1}) gives the equations of motion in the bulk, while
the boundary term  defines the boundary components of the HVF. Let us see what is the boundary contribution. From the previous discussion we have that the LHS of (\ref{HamEqs1}) can be obtained from (\ref{VarHT}), and, as a result, its boundary contribution is
\begin{equation}
\oint_{S_\Delta}\,\md^2\Theta\, (t\cdot V)\, Y_{\tilde{\pi}} = 
\oint_{S_\Delta}\md^2\Theta\, (X_{H2,\psi_I}Y_{\tilde{\pi}}-Y_{\psi_I}X_{H2,\tilde{\pi}})\, .
\end{equation}
It follows that
\begin{eqnarray}
 X_{H2, \tilde{\pi}} = 0 \ \ \ &\Rightarrow & \ \ \ \mathcal{L}_t\leftidx{^2}{\epsilon}=0\, ,\\
 X_{H2,\psi_I} = t\cdot V\ \ \ &\Rightarrow & \ \ \ \mathcal{L}_t\psi_I = t\cdot V\, ,
\end{eqnarray}
leading to the consistent boundary equations, as in the Case I. The main difference between the two cases is that in the first one we could obtain a well defined Hamiltonian, but we could not obtain the expression for the mass, while in the second one we could achieve both goals.   

As to gauge transformations and tangential diffeomorphisms,
all the results from Case I still hold, the only modificacion that has to be made is $\psi\to\psi_I$. Again, $U(1)$ transformation on WIH are gauge transformations, while the tangential diffeomorphisms have to be fixed. Let us now consider the case where we choose the other term to be dynamical.


\subsection{Case III}

In this case we shall take the opposite viewpoint to the previous case. That is, we shall have $(\psi_R,\tilde{\pi})$ as boundary degrees of freedom. This is the same choice that is made within the covariant phase space approach of, for example, \cite{AFK} and \cite{CRGV2016}. It turns out that we can obtain consistent boundary equations of motion, but again, we need to impose additional conditions and the boundary term in the corresponding well defined Hamiltonian vanishes.

In this case we rewrite the boundary term in the canonical action as
\begin{equation}\label{SboundarySP3}
 S_{\text{SD,3}}^{\Delta} =-\int \md t\oint_{S_\Delta} \, (\mathcal{L}_t\psi_R + t\cdot V ) \,\leftidx{^2}{\epsilon} \, .
\end{equation}
The symplectic structure again has a boundary term
\be\label{SymplStructSD3}
\Omega_3 =\int_\Sigma\md^3x\,\,\ed\tilde{P}^a_{\;AB}\dwedge \ed A_a^{\;AB} +  \oint_{S_\Delta}\md^2\Theta\, 
\,\ed\tilde{\pi}\dwedge\ed\psi_R\, ,
\ee
where the canonical variables on $S_\Delta$ are $(\psi_R,\tilde{\pi})$.
The canonical Hamiltonian is given by:
\begin{equation}
  H_{\text{SD,3}}=H_{\text{SD,1}}+\oint_{S_\Delta} \, (t\cdot V)\, \leftidx{^2}{\epsilon} \, .
\end{equation}
The boundary term should not contribute to the boundary equations of motion for $\psi_R$ or  $\tilde{\pi}$, nor to the mass of WIH, so we need to impose an additional condition
\be\label{Cond_tV}
\oint_{S_\Delta} \, (t\cdot V)\, \leftidx{^2}{\epsilon}=0 \, .
\ee
Then, 
$H_{\text{SD,3}}=H_{\text{SD,1}}$, and again the Hamiltonian weakly vanishes. 
From (\ref{VarH1}) and (\ref{Cond_tV}), we obtain 
\begin{equation}
 \delta H_{\text{SD,3}}\vert_{S_\Delta} = -\oint_{S_\Delta}\,  \kappa_{(t)}\, \delta\,\leftidx{^2}{\epsilon} 
 \, .\label{VarHamBound3}
\end{equation}
Under these restrictions, we can see that from
\begin{equation}
 \ed H_{\text{SD,3}}(Y) =\Omega_3 (Y, X_{\text{H3}})\, , \label{dHY}
\end{equation}
we can obtain the boundary components of the HVF, that lead to a consistent equations of motion on $\Delta$,
\be
 \mathcal{L}_t\leftidx{^2}{\epsilon}=0\, ,\ \ \ \ \mathcal{L}_t\psi_R = \kappa_{(t)} \, .
\ee

In the case of the Gauss constraint,
we look for a HVF $X_G$  that generates the infinitesimal gauge transformations, 
\begin{equation}
\ed G (Y) = \Omega_3 (Y, X_G)\, ,\label{GCVFEq3}
\end{equation}
where $Y$ is an arbitrary vector in $T\Gamma$. In this case the components of $X_G$ on the WIH are
\begin{equation}
 \delta_g\, \tilde{\pi} := X_{G,\tilde{\pi}} =  0
\end{equation}
\begin{equation}
 \delta_g\psi_R := X_{G,\psi_R} = -i\Lambda\, .
\end{equation}
The last expression represents the change in $\psi_R$ that is purely imaginary, so we must fix $\Lambda=0$ on WIH. Again, in this case the Gauss constraint does not generate gauge transformations on WIH. 
Similarly, boundary tangential diffeomorphisms have to be fixed.

\subsection{Case IV}

In this last case, we will examine the scenario where no boundary degrees of freedom are introduced and show that it also leads to a consistent description. In this case the Regge-Teitelboim type conditions hold and we can obtain the expression for the mass of WIH. Nevertheless, we will show that we need to impose some additional conditions on boundary variations and that the gauge transformations on WIH have to be completely fixed.  

In this case the boundary term in the canonical action is taken as
\be
S_{\text{SD,4}}^{\Delta}=-\int \md t \oint_{S_\Delta} (\kappa_{(t)} + t\cdot V)\,\leftidx{^2}{\epsilon} \, . 
\label{CanonicalAction4}
\ee
There are no boundary contributions to the symplectic structure
\be\label{SymplStructSD4}
\Omega_4 =\int_\Sigma\md^3x\,\,\ed\tilde{P}^a_{\;AB}\dwedge \ed A_a^{\;AB} \, ,
\ee
so that the variations of the allowed functionals cannot have a boundary term. 

The candidate for the canonical Hamiltonian is of the form
\begin{equation}
  H_{\text{SD,4}}=H_{\text{SD,1}}+\oint_{S_\Delta} \, (\kappa_{(t)}+ t\cdot V)\, \leftidx{^2}{\epsilon} \, .\label{Ham4}
\end{equation}
and its variation has a boundary term:
\begin{equation}
\delta H_{\text{SD,4}}\vert_{S_\Delta} = 
 \oint_{S_\Delta}\, \big[\, (\delta\kappa_{(t)} + \delta  (t\cdot V)\bigr]\, \leftidx{^2}{\epsilon}
 \, .\label{VarHamBound4}
\end{equation}
Since $\Omega_4$ does not have a boundary contribution, the boundary term (\ref{VarHamBound4}) must vanish or be canceled by a variaton of an appropriate counterterm, such that 
\be\label{VarHT3}
\delta (H_{\text{SD,4}}+H_{\text{CT,4}})\vert_{S_\Delta}=0 \, .
\ee
We have already seen that $H_{\text{CT,2}}$, given by (\ref{CT2}), cancels the first term in  (\ref{VarHamBound4}) and contributes to the correct expression for the mass of WIH. Any additional counter term that would annulate the second term in (\ref{VarHamBound4}), would change the expression for the mass.

In order to have a correct expression for a mass of the WIH , we need to impose an additional condition (\ref{Cond_tV}) in (\ref{Ham4}).
The final Hamiltonian, $H_4=H_{\text{SD,4}}+ H_{\text{CT,2}}$ is the same as in (\ref{TotalHamCT}), and, due to (\ref{Cond_tV}), its variation does not have a boundary term, that is consistent with the fact that the corresponding symplectic structure $\Omega_4$ does not have a boundary term.
We see that in this case, the dynamics is consistently described by the same total Hamiltonian as in the Case II, given in (\ref{TotalHamCT}). Nevertheless, we have an additional condition (\ref{Cond_tV}) and there are no boundary degrees of freedom on WIH. 

As to gauge transformations, 
since there are no boundary terms in the symplectic structure, from (\ref{SDGaussConstraint1}) and (\ref{G2}), it follows that $\Lambda =0$ on $\Delta$. Similarly, the boundary tangential diffeomorphisms are fixed on $S_\Delta$. 


\section{Discussion}

The purpose of the manuscript is to put forward a possibility of an interplay between a boundary symplectic structure and a boundary term in the Hamiltonian in diffeomorphism invariant theories defined in regions with boundary. Here we perform a canonical analysis of the theory of gravity in the first order formalism, in self-dual variables, defined in a region with a weakly isolated horizon as an internal boundary. We show that there are several choices of the boundary phase space and the corresponding well defined Hamiltonian. In this particular system we show that each of these choices lead to a consistent description of boundary dynamics, though in some of them  the corresponding Hamiltonian does not have a boundary term. We have also studied the residual gauge symmetries on the horizon. The results reported here represent the extension of the previous studies in two important directions. First, we are relaxing boundary conditions to WIH, in order to consider more general horizons. Next, we are exploring the freedom available in defining boundary phase spaces, that has not been taken previously into account. In that way we are extending some of the results obtained earlier, for strongly isolated horizons, that involved a particular choice of a boundary phase space.

One important issue that this results highlighted is that of the freedom (ambiguity?) involved in the definition of the theory on the boundary. As we have shown in detail, one can decide to introduce a potential on the boundary for the degrees of freedom that arise due to the IH boundary conditions. This is one possibility, but it is not a necessity. One can also take the Regge-Teitelboim perspective and have a well defined theory without any boundary contribution in the symplectic structure  nor in the variations of phase space functionals.
We analyzed this case in detail (our case IV).  We have seen that we get a consistent description and we {\it do} recover the (Smarr expression for the) mass of the horizon. We have also analyzed the cases in which one introduces a potential and have arrived at consistent formulations as well, that are not mutually equivalent. The introduction of potentials is the prevalent choice within the IH literature. The open question that remains is whether there is, or one can postulate, some physical principle that selects one of these possible choices. In the known case of electromagnetism, the introduction of a potential allows for a consistent action principle, and opens the door for gauge fields and gauge symmetries. Here we have indeed seen that (in one case) when a potential is introduced, one gains a $U(1)$ symmetry on the horizon. Whether this is the preferred choice remains to be seen.

Another important issue that one should comment on pertains to the possible quantization of the theory. In the original papers, based on spherically symmetric strongly isolated horizon with a fixed area \cite{ABCKprl,ACKclassical,ABKquantum}, an important feature of the resulting theory was that there was a degree of freedom that could be interpreted as a $U(1)$ Chern-Simons connection. This is the degree of freedom that is normally quantized, and has been shown to yield the quantum degrees of freedom responsible for black hole entropy.
In \cite{ENPprl,EngleNouiPerezPranzetti} a quantization of the same type of horizon with a $SU(2)$ Chern-Simons connection was also proposed.
In the present paper that involves, as we have already mentioned, several generalizations on different aspects, we  loose some of these features. For instance, we only recover the $U(1)$ gauge symmetry in some of the cases (I and II). However, we loose a direct (and obvious) interpretation of spatial diffeomorphism as gauge (as was the case in
\cite{ABCKprl,ACKclassical,ABKquantum}). We have shown in Appendix B that those original results can be extended to WIH for the non-rotating spherically symmetric case with a fixed area. Whether the original interpretation of quantum Chern-Simons degrees of freedom (with diffeo invariance) can be extended to generic WIH 
seems elusive and remains, in our opinion, an open issue.

In this manuscript we have restricted our attention to non-rotating horizons within the self-dual formulation. Generalizations to the rotating case and the more general Holst action are underway.
 
Finally, let us comment on another of the open questions pertaining the relationship between the canonical and covariant  approaches, especially between the corresponding boundary symplectic structures.   
Our results are a cautionary tale for the methods used in \cite{ENPprl,EngleNouiPerezPranzetti} to read from the covariant form the canonical symplectic structure directly used for quantization.
It would be interesting to further explore whether  the freedom that we found in the canonical description is also present in the covariant one. Investigations in this direction are ongoing.



\begin{acknowledgments}
This work was in part supported by DGAPA-UNAM IN114620 and CONACyT 0177840 grants, by CIC, UMSNH, and PRODEP 511-6/17-8204. 
\end{acknowledgments}

\appendix
\section{Potentials} \label{AppendixA}
  As we have seen in the main body of the manuscript, we can introduce the potentials $\psi_R$ and $\psi_I$ as degrees of freedom on the WIH, though they are just `parts' of the components of the self-dual connection, defined in  (\ref{real_psi}) and (\ref{imaginary_psi}). Here, we want to further explore in what extension these definitions determine the potentials.   
In order to obtain the  expansions of $\omega$ and $V$, we recall that in the comoving gauge one has \cite{CRVwih2}
\begin{align}
\omega =& -\kappa_{(\ell )} \mathbf{k} +\pi m+ \bar\pi\bar{m}\,, \label{WVUYdef2i} \\
V=&  -(\epsilon_{\text{NP}}-\bar{\epsilon}_{\text{NP}})\mathbf{k}+(\alpha-\bar{\beta})m+(\beta-\bar{\alpha})\bar{m}\,,   \label{WVUYdef2f}
\end{align}
where all forms are understood to be pulled back to $\Delta$ and $\pi$,  all equations are valid on $\Delta$, $\epsilon_{\text{NP}}$, $\alpha$ and $\beta$ are Newman-Penrose spin coefficients. Now, due to (\ref{real_psi}) and (\ref{imaginary_psi}) we have:
\begin{eqnarray}
\ell\cdot\omega = \ell\cdot \md\psi_R\ \ \Rightarrow\ \ \md\psi_R &=& - \kappa_{(\ell )} \mathbf{k} + c\, m+ \bar{c}\,\bar{m}\,,\label{Decomp3} \\
\ell\cdot V = \ell\cdot \md\psi_I\ \ \Rightarrow\ \ \md\psi_I &=&  -(\epsilon_{\text{NP}}-\bar{\epsilon}_{\text{NP}})\mathbf{k} + f\, m - \bar{f}\,\bar{m}\label{Decomp4} \,
\end{eqnarray}
where $c$ and $f$ are arbitrary functions on $\Delta$.

Comparing previous expansions, we see that we can make a decomposition
\begin{eqnarray}
\omega &=& \md\psi_R + \tilde\omega\, ,\label{Decomp1}\\
V &=& \md\psi_I + \tilde{V} \, ,\label{Decomp2} 
\end{eqnarray}
where 
\begin{eqnarray}
\tilde\omega &=& (\pi -c) m + (\bar{\pi}-\bar{c})\bar{m}\, ,\label{Decomp5}\\
\tilde{V} &=& (\alpha-\bar{\beta}-f)m -(\beta-\bar{\alpha}+\bar{f})\bar{m} \, .\label{Decomp6} 
\end{eqnarray}
WIH boundary conditions imply restrictions on $c$ y $f$. As shown, for example, in \cite{ChatterjeeGhosh}, on WIH we have
\begin{eqnarray}
\md\omega = 2 (\text{Im}\, \mathbf{\Psi_2})\,\leftidx{^2}{\epsilon}\, ,\\
\md V = 2 (\text{Re}\, \mathbf{\Psi_2})\,\leftidx{^2}{\epsilon}\, ,
\end{eqnarray}
where $\mathbf{\Psi_2}=C_{abcd}l^a m^b \bar{m}^c k^d$ is a complex scalar, the Newman-Penrose component of the Weyl tensor $C_{abcd}$.
For non-rotating WIHs, $\text{Im}\, \mathbf{\Psi_2}=0 $, so that $\md\omega = 0$, and particularly  $\md\tilde\omega = 0$ and $\md\tilde V \neq 0$, leading to conditions for $c$ and $f$. In contrast to the covariant phase space analysis, $\psi_R$ is not completely determined by $\kappa_{(l)}$, see, for example, \cite{ABL}. The same is true for $\psi_I$ and $V$.


\section{Spherically symmetric non-rotating WIH}\label{AppendixB}
Here, we shall restrict our study to spherically symmetric WIH, with fixed area $a_\Delta$ on the phase space. We shall consider all of our cases and see for which one we can obtain the Chern-Simons symplectic structure on the boundary.  This structure has been obtained previously in \cite{AFK}, for spherically symmetric non-rotating SIH with a preferred foliation. Their result can be extended to WIH, as shown in \cite{ChatterjeeGhosh}. For spherically symmetric non-rotating WIH, we have that $\text{Im}\,\Psi_2 \WIHeq 0$, that implies $\md\omega \WIHeq 0$, and also $m\cdot \omega \WIHeq 0$. One of the main relations that can be obtained in this case is  \cite{ChatterjeeGhosh}
\be
\md V \WIHeq -\frac{2\pi}{a_\Delta}\, \leftidx{^2}{\epsilon}\, .
\ee
Let us analyse what that implies in our four cases. We will also demonstrate that under these conditions in the first two cases boundary tangential diffeomorphisms are gauge symmetry of the theory. 

In the Case I, the boundary kinetic term is of the form 
\begin{equation}
 S_{\text{SD,1}}^{\Delta}=-\int \md t \oint_{S_\Delta} \mathcal{L}_{ t}\psi\,\leftidx{^2}{\epsilon}
 =-\int \md t \, \bigl[\oint_{S_\Delta} \mathcal{L}_{ t}\psi_R\,\leftidx{^2}{\epsilon}+ \oint_{S_\Delta} \mathcal{L}_{ t}\psi_I\,\leftidx{^2}{\epsilon}\bigr]
 \, ,
\end{equation}
The imaginary part can be rewritten as
\be\label{CSkinetic}
\oint_{S_\Delta}  \mathcal{L}_t\psi_I \,\leftidx{^2}{\epsilon} 
= -\frac{a_\Delta}{2\pi}\oint_{S_\Delta}  (t\cdot V) \,\md V =\frac{a_\Delta}{2\pi}
\oint_{S_\Delta} \md (t\cdot  V)\wedge V
= - \frac{a_\Delta}{2\pi}\oint_{S_\Delta} V\wedge \mathcal{L}_t V \, ,
\ee
where we have used that $\md (t\cdot V)=\mathcal{L}_t V -t\cdot\md V $ and $t\cdot\md V\sim \ell\cdot\ \leftidx{^2}{\epsilon}=0$. This is the $U(1)$ Chern-Simons boundary kinetic term in the canonical action.

Thus,  the boundary kinetic term is of the form
\be
S_{\text{SD,1}}^{\Delta}=-\int \md t \oint_{S_\Delta} \mathcal{L}_{ t}\psi_R\,\leftidx{^2}{\epsilon}+\frac{a_\Delta}{2\pi}\int \md t\oint_{S_\Delta} \, V\wedge\mathcal{L}_t V\, .
\ee
The boundary contribution to the symplectic structure is
\begin{equation}\label{OmegaSS}
 \Omega_1 (\delta_1,\delta_2)\vert_{\Delta}=\oint_{S_\Delta} (\delta_1\psi_R\,\delta_2\,\leftidx{^2}{\epsilon}-\delta_2\psi_R\,\delta_1\,\leftidx{^2}{\epsilon})
 -\frac{a_\Delta}{2\pi}\oint_{S_\Delta}\delta_1V\wedge\delta_2V\, .
\end{equation}
The first term in this expression vanishes, since due to spherical symmetry of $\kappa$, $\psi_R$ and $\delta\psi_R$ are also spherically symmetric, so that
\be
\oint_{S_\Delta} \delta_1\psi_R\,\delta_2\,\leftidx{^2}{\epsilon} = \delta_1\psi_R\, \oint_{S_\Delta} \delta_2\,\leftidx{^2}{\epsilon} = \delta_1\psi_R\, \delta_2 a_{\Delta}=0\, ,
\ee
since $a_\Delta$ is fixed in the phase space. 
The second term in $\Omega_1$ corresponds to the symplectic structure of the Chern-Simons theory on $S_\Delta$.

Finally, we only have
\begin{equation}\label{OmegaSS1}
 \Omega_1 (\delta_1,\delta_2)\vert_{\Delta}=
 -\frac{a_\Delta}{2\pi}\oint_{S_\Delta}\delta_1V\wedge\delta_2V\, .
\end{equation}

Let us analyze the dynamics on $\Delta$, starting from
\be
\delta H_{\text{SD,1}}|_\Delta = -\oint_{S_\Delta} t\cdot (\omega +V)\delta \,\leftidx{^2}{\epsilon}=   -\oint_{S_\Delta} (t\cdot \omega)\, \delta \,\leftidx{^2}{\epsilon}+ \frac{a_\Delta}{2\pi}\oint_{S_\Delta}\mathcal{L}_t V \wedge\delta V\, .
\ee
Again, the first term vanishes
\be
\oint_{S_\Delta} (t\cdot \omega)\, \delta \,\leftidx{^2}{\epsilon} = \kappa_{(t)} \delta a_\Delta =0\, .
\ee

The boundary components of the corresponding HVF are obtained from
\be
\delta H_{\text{SD,1}} = \Omega_1 (\delta , \delta_t)|_\Delta
\ee
leading to
\be
\delta_t V = \mathcal{L}_t V\, .
\ee

Let us now analyze the tangential diffeomorphisms on $S_\Delta$. From (\ref{deltaD}) we have 
\be
\delta D[N^a]\vert_{S_\Delta} 
=\oint_{S_\Delta} N\cdot  \delta [(\omega + V) \, \leftidx{^2}{\epsilon}] = 
\oint_{S_\Delta} \delta [(N\cdot V ) \, \leftidx{^2}{\epsilon}]\, .
\ee
For spherically symmetric WIH, $\pi =0$, so that $\omega = -\kappa_{(\ell )} \mathbf{k}$ and
$N\cdot\omega\sim m\cdot\omega =0$, implying $N\cdot\delta\omega =0$. For fixed $a_\Delta$, and due to (\ref{CSkinetic}), we further obtain that
\begin{eqnarray}
\delta D[N^a]\vert_{S_\Delta} 
&=&-\frac{a_\Delta}{2\pi}\oint_{S_\Delta}  [ \delta (N\cdot V)\, \md V + N\cdot V \, \delta(\md V)]\notag\\
{}&=& -\frac{a_\Delta}{2\pi}\oint_{S_\Delta} [(N\cdot\delta V)\,\md V - \md (N\cdot V)\wedge \delta V]\, .
\end{eqnarray}
Now, $(N\cdot\delta V)\,\md V=N\cdot (\delta V\wedge \md V)+\delta V\wedge N\cdot\md V$ and the first term in this expression vanishes since  $\delta V\wedge \md V$ is a 3-form on $S_\Delta$, so it follows that
\be \label{TanDiffSS}
\delta D[N^a]\vert_{S_\Delta}= -\frac{a_\Delta}{2\pi}  \oint_{S_\Delta} \delta V\wedge [N\cdot\md V + \md (N\cdot V)]
= -\frac{a_\Delta}{2\pi} \oint_{S_\Delta} \delta V\wedge\mathcal{L}_{\vec{N}}V\, . 
\ee
On the other hand,
\be
\delta D[N^a]=\Omega (\delta , \delta_{\vec{N}})\vert_{S_\Delta}\, .
\ee
Comparing (\ref{OmegaSS1}) with (\ref{TanDiffSS}), we see that, in this case, $D[N^a]$ generates tangential diffeomorphisms on $S_\Delta$, 
\be
\delta_{\vec{N}} V = \mathcal{L}_{\vec{N}}V\vert_{S_\Delta}\, .
\ee

In the Case II, for spherically symmetric WIH with fixed $a_\Delta$, using (\ref{CSkinetic}),  we obtain
\begin{equation}\label{SboundarySPSS}
 S_{\text{SD,2}}^{\Delta} =-\int \md t\oint_{S_\Delta} \, (\kappa_{(t)}\,  +  \mathcal{L}_t\psi_I ) \,\leftidx{^2}{\epsilon}=  \int \md t\, a_\Delta [-\kappa_{(t)}  +  \frac{1}{2\pi} \oint_{S_\Delta} \, V\wedge\mathcal{L}_t V ]\, .
\end{equation}
As in the previous case, we obtain a consistent dynamics on the horizon, 
$\delta_t V = \mathcal{L}_t V$. Again, $D[N^a]$ generates tangential diffeomorphisms on $S_\Delta$, 
$\delta_{\vec{N}} V = \mathcal{L}_{\vec{N}}V\vert_{S_\Delta}$.
We see that we recover the results of \cite{ABF}.

In the case III we have
\begin{equation}\label{SboundarySP4SS}
 S_{\text{SD,4}}^{\Delta} =-\int \md t\oint_{S_\Delta} \, (\mathcal{L}_t\psi_R + t\cdot V ) \,\leftidx{^2}{\epsilon} \, .
\end{equation}
As we have seen in the Case I, the first term does not contribute to the symplectic structure and the second one has to vanish.
In both cases there is no boundary contribution to the symplectic structure. 

Finally, in case IV there is no boundary contribution to the symplectic structure.


\end{document}